\documentclass[10pt]{amsart}   
\usepackage{amssymb,amsmath,amscd,amsfonts,amsthm,mathrsfs,curves}
\usepackage[active]{srcltx} 
\usepackage[all]{xy}  
\usepackage{color}

\newtheorem{theorem}{Theorem}[section] 
 
\newtheorem{proposition}[theorem]{Proposition}

\theoremstyle{definition}

\theoremstyle{remark}  
\newtheorem{rem}[theorem]{Remark}  
  
\newcommand{\Bc}{\mathcal{B}}  
\newcommand{\Cc}{\mathcal{C}}  
\newcommand{\Dc}{\mathcal{D}} 
\newcommand{\Ec}{\mathcal{E}}

\newcommand{\Hc}{\mathcal{H}}  
\newcommand{\Ic}{\mathcal{I}}  
\newcommand{\Jc}{\mathcal{J}}  
 
\newcommand{\Mcc}{\mathcal{M}}

\newcommand{\Vc}{\mathcal{V}}

\newcommand{\C}{\mathbb{C}} 
\newcommand{\Hbb}{\mathbb{H}} 
\newcommand{\R}{\mathbb{R}}  
\newcommand{\N}{\mathbb{N}}  

\newcommand{\D}{\mathbb{D}}

\def\build#1_#2^#3{\mathrel{\mathop{\kern 0pt#1}\limits_{#2}^{#3}}} 
 
\newcommand{\un}{1\hskip-0.6ex {\mathsf{ I} }}

\newcommand{\donne}{\mapsto}
\newcommand{\dans}{\longrightarrow}

\newcommand{\rL}{{\rm L}}

\begin{document}  
\title[Oscillating potentials]
{On Schr\"odinger and Dirac operators with an
oscillating potential.}
\author{Thierry Jecko\\
{\tiny with contributions by} Aiman Mbarek.}
\address{AGM, UMR 8088 du CNRS, Universit\'e de Cergy Pontoise, Site de Saint Martin,  
2, rue Adolphe Chauvin, F-95302 Cergy-Pontoise cedex, France.}
\email{thierry.jecko@math.cnrs.fr\, .}
\date{\today}   
\begin{abstract}
We review some results on the spectral theory of Schr\"odinger and Dirac operators. We focus on two aspects: the 
existence of embbedded eigenvalues in the essential spectrum and the limiting absorption principle. They both 
are important for Physics, in general, and for Scattering theory, in particular. We chose to include a special form of 
oscillations in the potential of the considered operators to illustrate the diversity of behaviours that can 
exist in the two selected topics. Concerning the limiting absorption principle, we discuss several methods 
to prove it. To known, old or recent results, we 
added some unpublished results from Mbarek's Phd Thesis. 
\end{abstract}   
\maketitle   

\section{Introduction.} \label{s:intro}
\setcounter{equation}{0}

The purpose of this paper is to review some results on the spectral theory of (mainly) Schr\"odinger and Dirac operators 
containing some oscillating part in their potential. This is of course a vast subject that we cannot reasonably 
cover in a short paper. We shall restrict ourselves to two phenomena, the existence of embedded eigenvalue in the continuous spectrum and 
the so-called limiting absorption principle (LAP) (see \eqref{eq:tal-A} and~\eqref{eq:tal-Q} below). \\
The first topic was historically considered as a kind of anomaly, but it turns out that it is not so marginal in Physics, as one would first think. Furthermore, its understanding and its influence on Scattering theory constitute 
a mathematical challenge. The second topic is a cornerstone of the stationary version of Scattering theory 
(cf. \cite{rs3,rs4,y}). The LAP directly enters in representation formulae for Scattering operators and matrices 
(cf. \cite{y}). \\
The interest in oscillating potentials comes from the fact that, for many questions in spectral theory, such potentials 
require an different treatment to classical ones. 
Furthermore, in some cases, they even constitute an exception to the behaviour that is observed or even proved for 
a large class of models. Many mathematical works, mainly in the seventies and the eighties, confirm these facts, 
see \cite{w} and references therein. \\
In this review, we want to illustrate this ``anomalous'' behaviour of oscillating potentials. To this end, it actually 
suffices to consider a quite narrow class of such potentials. We shall indeed see, for instance, that some particular 
potentials of the form \eqref{eq:W} below, can produce an embedded eigenvalue in the continuous spectrum of a Schr\"odinger operator 
while this is forbidden for a large class of potentials. We shall also see that known methods to get the LAP are simply 
inapplicable for some Schr\"odinger operators containing such an oscillating part in their potentials. Nethertheless the 
LAP does hold true for many of them. It is however expected that the LAP breaks down near certain energy for some 
cases of this kind. We chose to only consider Schr\"odinger and Dirac operators acting on functions of a continuous 
variable in $\R^d$, $d\in\N^\ast$, but we mention that there exist recent results of this kind on discrete 
Schr\"odinger operators (cf. \cite{man}). There are also interesting works concerning oscillating potentials that 
treat related subjects, like the nature of the essential spectrum or the wave operators (see \cite{ki,lns,rem,w}). We decided  
not to elaborate on this. \\
Among the quite big litterature on the subject, we made a selection of works, that we considered as relevant for the 
above illustration. We did not try at all to provide an exhaustive exposition of this subject. We picked up some recent 
results, some of them being unpublished. In particular, we present here unpublished results obtained by A. Mbarek in 
his Phd Thesis (cf. \cite{mb}), namely Proposition~\ref{p:w-v-n-dirac-dim-3}, Theorem~\ref{th:tal-A-Wigner-Dirac}, 
Theorem~\ref{th:tal-Q-m}, and Proposition~\ref{p:oscillations-energy1-bis} below, and provide for each a 
(sketch of) proof. \\
We also decided not to discuss open questions. Even in the chosen, quite narrow area of study, there are many interesting 
unsolved questions. This would considerably increase the length of the paper. Nevertheless, for each result, 
the reader could ask himself what happens if one assumption is removed (or weakend). He would probably 
get an interesting problem since our understanding of oscillations in the present context is quite limited. 
Let us just mention one question: While the paper provides situations for which the LAP is valid, 
is there in the selected class of oscillating potentials an example for which 
the LAP breaks down?

The paper is organized as follows. Section~\ref{s:embedded} is devoted to the question of embbedded eigenvalues. 
In Subsection~\ref{ss:existence}, we provide examples of the chosen potential class for which there exists 
(at least) an embbedded eigenvalue. This is done for Schr\"odinger, Dirac and Klein-Gordon operators. 
In Subsection~\ref{ss:dense}, we present a situation in which so many 
embbedded eigenvalues are produced that the point spectrum is dense in some part of the essential spectrum. 
Subsection~\ref{ss:absence} states, for the considered potential class, results showing the absence of embbedded 
eigenvalues in a large part of the essential spectrum. Among the class, this gives us a quite precise description of 
those oscillating potentials that produce embbedded eigenvalues and also where these eigenvalues 
are located. Section~\ref{s:tal} treats the LAP. Subsection~\ref{ss:ode} focuses on $1$-dimensional technics of 
ordinary differential equations that apply to the $1$-dimensional and radial cases. 
To cover more general situations, powerful methods are presented in an abstract way in Subsection~\ref{ss:putnam}. In particular Mourre's commutator method is briefly reviewed and two versions of the recent ``local Putnam-Lavine theory'' are described in details. In Subsection~\ref{ss:almost-short-range}, we apply the first version to Schr\"odinger and Dirac operators with almost short-range oscillations in the potential. The more complicated case of long-range oscillations for Schr\"odinger operators is treated in Subsection~\ref{ss:weak-decaying} with the second version of the ``local Putnam-Lavine theory''. We ploted in Figure~\ref{dessin:1} the actual state of validity of the LAP for 
the considered Schr\"odinger operators. 

The rest of this Introduction is devoted to the presentation of the Hamiltonians we study and to a precise 
statement of their LAP. To this end, we need to introduce some notation.
Let $d\in\N^\ast$. We denote by $\langle \cdot ,\cdot \rangle$ and $\|\cdot\|$ the right
linear scalar product and the norm in $\rL^2(\R^d)$, the space of squared integrable, complex
functions on $\R^d$. We also denote by $\|\cdot\|$ the norm of bounded operators on $\rL^2(\R^d)$.
Writing $x=(x_1; \cdots ; x_d)$ the variable in $\R^d$, we set
\[|x| \ :=\ \biggl(\sum_{j=1}^d\, x_j^2\biggr)^{1/2}
\hspace{.4cm}\mbox{and}\hspace{.4cm}\langle x\rangle \ :=\ \biggl(1\, +\, \sum_{j=1}^d\, x_j^2\biggr)^{1/2}\, .\]
Let $Q_j$ the multiplication operator in $\rL^2(\R^d)$
by $x_j$ and $P_j$ the self-adjoint realization of $-i\partial _{x_j}$ in $\rL^2(\R^d)$. We set $Q=(Q_1; \cdots ; Q_d)^T$
and $P=(P_1; \cdots ; P_d)^T$, where $T$ denotes the transposition. \\
Several times, we shall use the following compacity result in conjonction with 
the resolvent form of Weyl's theorem on the essential spectrum (see \cite{rs4}, Section XIII.4). If $V$ is a complex-valued, bounded function on $\R^d$, that tends to $0$ at infinity, and $f$ is a real, non-constant polynomial function, then $V(Q)(f(P)+i)^{-1}$ is compact. Thus, by the resolvent formula, the difference of the resolvent of $f(P)+V(Q)$ and of the resolvent of $f(P)$ is compact. In particular, $f(P)+V(Q)$ is self-adjoint on the domain of $f(P)$ and, by Weyl's theorem, its essential spectrum $\sigma _{\rm ess}(f(P)+V(Q))$ coincides 
with the one of $f(P)$, which is also the spectrum $\sigma (f(P))$ of $f(P)$. All this still holds true if $V$ (resp. $f$) has values in the set of squared matrices with complex entries 
(resp. the set of self-adjoint, squared matrices with complex entries). \\
Let
\[H_0\ =\ |P|^2\ :=\ \sum_{j=1}^d\, P_j^2\ =\ P^T\cdot P\]
be the self-adjoint realization of the nonnegative Laplace operator
$-\Delta$ in $\rL^2(\R^d)$. Its domain is the Sobolev space $\Hc^2(\R^d)$. 
We shall mainly consider Schr\"odinger operators $H=H_0+V(Q)$, where $V(Q)$ is the
multiplication operator by a real valued function $V$ on $\R^d$. We shall focus on the case where $V$ contains 
a (several) oscillating part(s) of the form 
\begin{equation}\label{eq:W}
W_{\alpha\beta}(x)\ =\ w\bigl(1 - \kappa (|x|)\bigr)|x|^{-\beta}\sin (k|x|^\alpha)
\end{equation}
with real, nonzero $w$ and $k$, $\alpha >0$, $\beta>0$ and $\kappa :\R\dans\R$ a smooth, compactly supported 
function that is $1$ near $0$. The parameter $w$ controls the strength of the potential. We chose to focus on 
the behaviour of the potential at infinity and not to consider a possible singularity at zero of the function $x\donne |x|^{-\beta}\sin (k|x|^\alpha)$. 
So we introduced the cut-off function $\kappa$ to exclude such possibility. 
The potential is short-range if $\beta>1$, else long-range. Finaly the parameter $\alpha $ controls the speed of oscillations at infinity. 
Of course, one can replace the sinus function above by a cosinus function. \\
We also allow $V$ to contain a {\em short-range} part $V_{sr}$ and a {\em long-range} part $V_{lr}$ 
satisfying the following conditions: there exist $\rho _{sr}, \rho _{lr}, \rho _{lr}'\in ]0; 1]$ such 
that the real functions $\R^d\ni x\donne\langle x\rangle^{1+\rho _{sr}}V_{sr}(x)$ and 
$\R^d\ni x\donne\langle x\rangle^{\rho _{lr}}V_{lr}(x)$ are bounded, and such that the distribution 
$\langle x\rangle^{\rho _{lr}'}x\cdot\nabla V_{lr}(x)$ co\"\i ncide with a bounded function. In almost 
all considered cases, $H$ will be self-adjoint with domain $\Hc^2(\R^d)$ and, due to Weyl's theorem 
and the fact that $V$ tends to $0$ at infinity, its essential spectrum will be the 
one of $H_0$, namely $[0; +\infty[$.\\
For a large class of potentials $V$, for compact intervals $\Ic\subset ]0; +\infty[$, one can show the following 
LAP (see for instance \cite{abg}), that is the bound, for $s>1/2$,   
\begin{equation}\label{eq:tal-A}
\sup_{\Re z\in\Ic,\atop\Im z\neq 0}\bigl\|\langle A\rangle^{-s}(H-z)^{-1}\langle A\rangle^{-s}\bigr\|\ <\ +\infty
\, ,
\end{equation}
where $A$ is the self-adjoint realization in $\rL^2(\R^d)$ of the differential operator 
$2^{-1}(P\cdot Q+Q\cdot P)$. While the resolvent $(H-z)^{-1}$ blows up as bounded operator on $\rL^2(\R^d)$ as 
the spectral parameter $z$ approaches the spectrum in $\Ic$, the introduction of weights $\langle A\rangle^{-s}$ 
on both sides with $s>1/2$ ``absorbs'' this blow up. Sometimes it is convenient to have the same result with 
the generator of dilations $A$ replaced by the position operator $Q$, namely, for $s>1/2$,  
\begin{equation}\label{eq:tal-Q}
\sup_{\Re z\in\Ic ,\atop \Im z\neq 0}\bigl\|\langle Q\rangle^{-s}(H-z)^{-1}\langle Q\rangle^{-s}\bigr\|\ <\ +\infty
\, ,
\end{equation}
which is actually a consequence of \eqref{eq:tal-A}. 
We note that \eqref{eq:tal-A} (resp. \eqref{eq:tal-Q}) cannot hold true for some $s>1/2$ if $H$ has an eigenvector 
in the domain of $\langle A\rangle^{s}$ (resp. $\langle Q\rangle^{s}$) associated to an eigenvalue in $\Ic$. Therefore, 
it is natural to seek for a LAP on an interval $\Ic$ that avoids the point spectrum of $H$. Nethertheless, it is 
sometimes possible to allow point spectrum in $\Ic$ provided that one lets the resolvent act on the orthogonal 
complement of the spectral subspace associated to this point spectrum part. For simplicity, we chose not to discuss this issue. 

Let us introduce now the Dirac operators we shall consider. Let 
\[\sigma _1\ =\ \left(
\begin{array}{cc}
0&1\\
1&0
\end{array}\right)\, ,\hspace{.4cm}
\sigma _2\ =\ \left(
\begin{array}{cc}
0&-i\\
i&0
\end{array}\right)\, ,
\hspace{.4cm}
\sigma _3\ =\ \left(
\begin{array}{cc}
1&0\\
0&-1
\end{array}\right)\, ,
\]
be the Pauli $2\times 2$ matrices and, denoting by $\un _2$ the $2\times 2$ identity matrix, let 
\[\beta\ =\ \left(
\begin{array}{cc}
\un_2&0\\
0&-\un_2
\end{array}\right)\, ,\hspace{.4cm}
\alpha _j\ =\ \left(
\begin{array}{cc}
0&\sigma _j\\
\sigma _j&0
\end{array}\right)\, ,
\]
for $j\in \{1; 2; 3\}$, be the Dirac $4\times 4$ matrices. Choosing the speed of light to be $1$, 
the free Dirac operator $\D _0$ in $\R^3$ with mass $m\in\R^+$ is the self-adjoint realization in 
$\rL^2(\R^3; \C^4)\sim\rL^2(\R^3)\otimes\C^4$ of the differential operator 
\begin{equation}\label{eq:free-dirac}
\D _0\ =\ \underline{\alpha}\cdot P\, +\, m\, \beta := \sum _{j=1}^3\alpha _j\, P_j\, +\, m\, \beta\, ,
\end{equation}
where $\underline{\alpha}$ is the transposed vector $(\alpha _1; \alpha_2; \alpha _3)^T$ of three $4\times 4$ matrices. 
Its domain is the Sobolev space $\Hc^1(\R^3; \C^4)\sim\Hc^1(\R^3)\otimes\C^4$. We shall consider (electromagnetic) 
external fields that is functions $V$ that are defined on $\R^3$ with values in the space of self-adjoint $4\times 4$ matrices and 
tend to $0$ at infinity. In particular, the full Dirac operator $\D=\D _0+V(Q)$ will always be self-adjoint with domain 
$\Hc^1(\R^3; \C^4)$ and the essential spectrum of $\D$ will be the one of $\D_0$, that is $]-\infty ; -m]\cup 
[m; +\infty[$, by Weyl's theorem. In one dimension, on the line or the half-line with boundary conditions, it is 
standard (cf. \cite{sc}) to view the free Dirac operator with mass $m$ as the operator 
\begin{equation}\label{eq:free-dirac-dim1}
\D_0^1\ :=\ \sigma _2\, P\, +\, m\, \sigma _3\, ,
\end{equation}
acting in $\rL^2(\R; \C^2)$ or in $\rL^2(]0; +\infty[; \C^2)$ and the full Dirac operator $\D^1$ given by 
$\D_0^1+V(Q)$ where the electromagnetic potential $V$ takes its values in the space of (self-adjoint) $2\times 2$ matrices. 
This is a natural definition since the $3$-dimensional free Dirac operator $\D_0$ reduces to a certain direct 
sum, indexed by triplets $\rho =(j_\rho; m_\rho; \kappa _\rho)$ of numbers, of operators $\D_0^1+\kappa_\rho Q^{-1}$ 
acting on $\rL^2(]0; +\infty[; dr)\otimes h_\rho$, for some two dimensional spaces $h_\rho$ (cf. \cite{th}, p. 128). \\
It is also of interest to prove a LAP for the Dirac operator. For a large class of external potential $V$ and $m>0$, 
for compact intervals $\Ic\in ]-\infty; -m[\cup ]m; +\infty[$ that does not intersect the point spectrum of $\D$, 
one can show, for $s>1/2$, 
\begin{equation}\label{eq:tal-A-dirac}
\sup_{\Re z\in\Ic,\atop\Im z\neq 0}\bigl\|\langle A'\rangle^{-s}(\D -z)^{-1}\langle A'\rangle^{-s}\bigr\|\ <\ +\infty
\, ,
\end{equation}
where $A'$ is a relatively explicit, self-adjoint operator (cf. \cite{bmp}). This result often implies the LAP with weights 
$\langle Q\rangle$, namely, for $s>1/2$, 
\begin{equation}\label{eq:tal-Q-dirac}
\sup_{\Re z\in\Ic,\atop\Im z\neq 0}\bigl\|\langle Q\rangle^{-s}(\D -z)^{-1}\langle Q\rangle^{-s}\bigr\|\ <\ +\infty
\, .
\end{equation}

{\bf Aknowledgements:}
The author thanks A. Mbarek for allowing the inclusion in the text of several unpublished results of his Phd Thesis 
\cite{mb} (2017). He is also grateful to I. Herbst for pointing out to him White's paper \cite{w}. The author also 
thanks J. Faupin, M. Mantoiu, and V. Nistor for the invitation to the Conference at Metz in May 2017, where some results of the present paper 
were presented. \\
{\em Dedicated in alphabetic order to} Ca., Ch., Ch., Ch., Cl., and K.

\section{Embedded eigenvalues in the continuous spectrum.} 
\label{s:embedded}
\setcounter{equation}{0}

In this section, we discuss the question of the existence of embbedded eigenvalues in the essential spectrum 
of Schr\"odinger and Dirac operators. It is easy to produce such embbedded eigenvalues using a direct 
sum of operators, one of them having (say) continuous spectrum and others having point (or discrete) spectrum. 
Furthermore, when an Hamiltonian has such embbedded eigenvalues, it can always be rewritten as a direct 
sum of this kind. The interest for oscillating potentials is that they provide examples of this phenomenon 
while a direct sum structure of the above kind is a priori not apparent. 

\subsection{Existence of embedded eigenvalues.} 
\label{ss:existence}

In this subsection, we show that potentials containing an oscillating part of the form \eqref{eq:W} can produce an 
embedded eigenvalue in the continuous spectrum of Schr\"odinger, Klein-Gordon, and Dirac operators. 

We start with the example given by Wigner and Von Neumann in \cite{vw}. See \cite{rs4}, p. 223, for an heuristic 
and a proof. 
\begin{proposition}\label{p:w-v-n-dim-1}
Let $f, g, V : \R\dans\R$ be defined by $g(x)=2x-\sin (2x)$, $f(x)=(1+g(x)^2)^{-1}\sin (x)$ and 
\begin{equation}\label{eq:w-v-n}
V(x)\ =\ -16\cdot\frac{g(x)\cdot\sin (2x)}{1+g(x)^2}\, -\, 
32\bigl(1\, -\, 3g(x)^2\bigr)\cdot\frac{\sin^3(x)}{\bigl(1+g(x)^2\bigr)^{2}}\, .
\end{equation}
Then $f, f'\in\rL^2(\R)$ and $-f''+Vf=f$. 
\end{proposition}  
We point out that, in the proof of Propositions~\ref{p:w-v-n-dim-1} given in \cite{rs4}, the bound state $f$ is first constructed and 
the potential $V$ is then defined by $(f''+f)/f$. Since $f$ has zeroes, one has to choose $g$ carefully to avoid singularities 
in $V$. There is a $3$-dimensional version of this Proposition, namely 
\begin{proposition}\label{p:w-v-n-dim-3}
Let $f, g, W : \R\dans\R$ be defined by $g(x)=2|x|-\sin (2|x|)$, $f(x)=(1+g(x)^2)^{-1}|x|^{-1}\sin (|x|)$ and 
$W(x)=V(|x|)$, where $V$ is given by \eqref{eq:w-v-n}. 
Then $f, \nabla f\in\rL^2(\R^3)$ and $-\Delta f+Wf=f$. 
\end{proposition}  
In both cases, $1$ is an eigenvalue of the Schr\"odinger operator. We note that there exists $C>0$ such that, for 
all $t\in\R$, $|V(t)|\leq C\langle t\rangle^{-1}$. Thus, by Weyl's theorem, both Schr\"odinger operators has the 
same essential spectrum as the Laplace operator $-\Delta$, namely $[0; +\infty[$. Thus $1$ is an embedded eigenvalue 
in the continuous spectrum. 

The above idea can be recycled to get an embedded eigenvalue for the one-dimensional Klein-Gordon operator with mass 
$m\geq 0$ which is the self-adjoint realization of $\sqrt{P^2+m^2}-m+V(Q)$. This has been done 
recently in \cite{ls} in the massive and the massless cases. In particular, in the massive case, one obtain the 
\begin{proposition}\label{p:w-v-n-klein-gordon-dim-1} \cite{ls}. 
Let $m>0$ and $\lambda :=\sqrt{1+m^2}-m>0$. Let $f, g, h, k : \R\dans\R$ be defined by $g(x)=2x-\sin (2x)$, 
$h(x)=(1+g(x)^2)^{-1}$, 
\[k\ =\ \bigl(\sqrt{(P+1)^2+m^2}\, +\, \sqrt{(P-1)^2+m^2}\bigr)h\, ,\]
and $f(x)=k(x)\sin (x)$. For all $x\in\R$, $V(x)=\lambda - f(x)^{-1}((\sqrt{P^2+m^2}-m)f)(x)$ is well-defined in $\R$ 
and the function $V$ is, in absolute value, bounded above by some $C\langle\cdot\rangle^{-1}$. Furthermore, 
$f, f'\in\rL^2(\R)$ and $(\sqrt{P^2+m^2}-m)f+Vf=\lambda f$. 
\end{proposition}  

Due to the non-local structure of the kinetic energy, the arguments are more involved than for the Schr\"odinger case. 
This appears in particular when one wants to ensure that the zeroes of $f$ do not produce singularities in $V$. 

It turns out that one can replace $\sin (2x)$ above by $\sin (kx)$ with an appropriate $k>0$ to get an embedded eigenvalue at a 
given value in the continuous spectrum. This has also been done for the Dirac operator in $\R^3$ with a radial 
electromagnetic potential $\D=\D_0+V(Q)$ (see \eqref{eq:free-dirac}). In particular, one has the 
\begin{proposition}\label{p:w-v-n-dirac-dim-3} \cite{mb}. 
Let $m\geq 0$ and $\lambda\in\R$ such that $|\lambda |>m$. Then one can find a radial, bounded function $V$ with values in the set of self-adjoint $4\times 4$ matrices such that $V$ tends to $0$ at infinity, $\lambda$ is an eigenvalue of $\D=\D_0+V(Q)$, $\D$ is self-adjoint on the domain of $\D_0$, namely $\Hc^1(\R^3; \C^4)$, and $\sigma _{\rm ess}(\D)=\sigma (\D_0)$. 
\end{proposition}
One can follow here the heuristic used in \cite{rs4}, p. 223, to prove Proposition~\ref{p:w-v-n-dim-1}. Let 
us sketch this and the proof of Proposition~\ref{p:w-v-n-dirac-dim-3}. 

First of all, a radial potential is, according to \cite{th} p. 128 and denoting by $\un _4$ the $4\times 4$ identity 
matrix, of the form 
\begin{equation}\label{eq:pot-radial-dirac}
V(x)\ =\ \phi _{el}(|x|)\, \un _4\, +\, \phi _{sc}(|x|)\, \beta \, +\, i\phi _{am}(|x|)\, \beta\, \underline{\alpha}
\cdot x|x|^{-1}
\end{equation}
with real valued functions $\phi _{el}$, $\phi _{sc}$, and $\phi _{am}$, on $]0; +\infty[$. By Theorem 4.14 in 
\cite{th}, p. 128, it turns out that the radial Dirac operator $\D=\D_0+V(Q)$ is unitary equivalent to a direct sum 
$\oplus _{\rho\in\Lambda}\D_\rho$ of operators $\D_\rho$ acting in $\rL^2(]0; +\infty[; dr)\otimes h_\rho$, for 
some two dimensional space $h_\rho$. The parameters $\rho$ are actually triplets of eigenvalues of $3$ appropriate 
operators that commute with $\D$. Taking an appropriate basis $\Bc_\rho$ of each $h_\rho$, $\D_\rho$ acts as 
the $1$-dimensional Dirac operator (in the variable $r$) 
\[\D_\rho\ =\ \sigma _2\, P_r\, +\, \bigl(m\, +\, \phi _{sc}(Q_r)\bigr)\, \sigma_3\, +\,
\Bigl(\frac{\kappa_\rho}{Q_r}\, +\, \phi _{am}(Q_r)\Bigr)\, \sigma _1\, +\, \phi _{el}(Q_r)\, \un _2\, ,
\]
where $\kappa _{\rho}$ is one of the eigenvalues in $\rho$ and is a nonzero integer. This can also be written as 
\[\D_\rho\ =\ \left(
\begin{array}{cc}
m+\phi _{sc}(Q_r)+\phi _{el}(Q_r)&-\frac{d}{dr}+\frac{\kappa_\rho}{Q_r}+\phi _{am}(Q_r)\\
\frac{d}{dr}+\frac{\kappa_\rho}{Q_r}+\phi _{am}(Q_r)&-m-\phi _{sc}(Q_r)+\phi _{el}(Q_r)
\end{array}
\right)\, .\]
Now we look at the Dirac equation $\D u=\lambda u$ in the new representation, namely in: 
\begin{equation}\label{eq:direct-sum}
 \bigoplus _{\rho\in\Lambda}\ \Bigl(\rL^2(]0; +\infty[; dr)\otimes h_\rho\Bigr)\, .
\end{equation}
We choose some arbitrary $\rho_0\in\Lambda$. We shall construct a radial potential $V$ satisfying the conditions in Proposition~\ref{p:w-v-n-dirac-dim-3} such that 
there is a bound state $f=(f_\rho )_{\rho\in\Lambda}$ at energy $\lambda$ such that $f_\rho =0$, if $\rho\neq\rho_0$, and $f_{\rho _0}$ is an appropriate nonzero vector in $\rL^2(]0; +\infty [; dr)\otimes h_{\rho_0}$. Using the previously mentioned basis $\Bc_{\rho_0}$ of 
$h_{\rho _0}$, the vector $f_{\rho _0}$ is represented by some $(f_1; f_2)^T\in\rL^2(]0; +\infty[; \C^2)$. 
For simplicity, we denote $\kappa_{\rho _0}$ by $\kappa$. \\
Since the potential should be small at infinity, we guess the form of $(f_1; f_2)^T$ by solving the 
system 
\[\left(
\begin{array}{cc}
m&-\frac{d}{dr}\\
\frac{d}{dr}&-m
\end{array}
\right)\cdot \left(
\begin{array}{c}
g_1\\
g_2
\end{array}
\right)\ =\ \lambda \left(
\begin{array}{c}
g_1\\
g_2
\end{array}
\right)\, .\]
The solutions of this system are, for $(g_1^0; g_2^0)\in\R^2$, the functions 
\[r\ \donne\ \left(
\begin{array}{c}
g_1\\
g_2
\end{array}
\right)(r)\ =\ \exp (rM)\cdot\left(
\begin{array}{c}
g_1^0\\
g_2^0
\end{array}
\right)\, ,\hspace{.4cm}\mbox{with}\hspace{.4cm}M\ =\ \left(
\begin{array}{cc}
0&k_+\\
k_-&0
\end{array}
\right)\]
and $k_\pm = m\pm\lambda$. Note that the eigenvalues of $M$ are $\pm i\sqrt{-k_+k_-}$. Thus $\exp (rM)$ carries 
oscillating terms. These terms will be responsible for oscillations in $V$. 
Note also that the function $]0; +\infty[\ni r\donne\exp (rM)$ is bounded. \\
For appropriate real smooth functions $u_1$ and $u_2$, the bound state will be given by 
\[\left(
\begin{array}{c}
f_1\\
f_2
\end{array}
\right)(r)\ =\ \exp (rM)\cdot\left(
\begin{array}{c}
u_1(r)\\
u_2(r)
\end{array}
\right)\, .\]
Now, the equation 
\begin{equation}\label{eq:dirac-eq-h-rho}
\D_\rho\cdot \left(
\begin{array}{c}
f_1\\
f_2
\end{array}
\right)\ =\ \lambda \left(
\begin{array}{c}
f_1\\
f_2
\end{array}
\right)
\end{equation}
is equivalent to 
\begin{equation}\label{eq:cond-V}
 i\sigma_2\exp (rM)\frac{d}{dr}\left(
\begin{array}{c}
u_1\\
u_2
\end{array}
\right)\ =\ V(r)\exp (rM)\left(
\begin{array}{c}
u_1\\
u_2
\end{array}
\right)\, .
\end{equation}
Setting 
\[\left(
\begin{array}{c}
v_1\\
v_2
\end{array}
\right)\ =\ \exp (rM)\left(
\begin{array}{c}
u_1\\
u_2
\end{array}
\right)\hspace{.4cm}\mbox{and}\hspace{.4cm}\left(
\begin{array}{c}
w_1\\
w_2
\end{array}
\right)\ =\ i\sigma_2\exp (rM)\frac{d}{dr}\left(
\begin{array}{c}
u_1\\
u_2
\end{array}
\right)\, ,\]
and requiring that $u_1^2+u_2^2$ does not vanish, we can solve \eqref{eq:cond-V} for $V$ and get 
\begin{equation}\label{eq:comp-V-1}
\phi _{sc}\ =\ \frac{v_1w_1-v_2w_2}{v_1^2+v_2^2}\, +\, \frac{v_2^2-v_1^2}{v_1^2+v_2^2}\cdot\phi _{el}
\end{equation}
and 
\begin{equation}\label{eq:comp-V-2}
\frac{\kappa}{r}\, +\, \phi _{am}\ =\ 
\frac{v_1w_2+v_2w_1}{v_1^2+v_2^2}\, -\, \frac{2v_1v_2}{v_1^2+v_2^2}\cdot\phi _{el}\, ,
\end{equation}
where $\phi _{el}$ is arbitrary. Now, as we shall see in the sketch of the proof of Proposition~\ref{p:w-v-n-dirac-dim-3} 
below, one can choose the functions $u_1$, $u_2$, and $\phi _{el}$ such that this $V$ satisfies 
the requirement of Proposition~\ref{p:w-v-n-dirac-dim-3}, such that $\lambda$ is an eigenvalue of $\D$, and such 
that an eigenvector $f$ of $\D$ associated to the eigenvalue $\lambda$ is given in the direct sum \eqref{eq:direct-sum} by $f_\rho=0$, if $\rho\neq\rho _0$, and $f_{\rho _0}=(f_1; f_2)^T$. 

The massive threshold case $|\lambda |=m>0$ has also been treated in Mbarek's Phd Thesis \cite{mb}, in a similar way. 
Compared to the Schr\"odinger case, the arguments are easier. This is due to the facts that we have two degrees of 
freedom in \eqref{eq:dirac-eq-h-rho} and that the Dirac operator is of order $1$. In particular, it is quite easy 
to ensure that the divisions in \eqref{eq:comp-V-1} and~\eqref{eq:comp-V-2} do not produce singularities in the 
components $\phi _{sc}$, $\phi _{am}$, and $\phi _{el}$, of $V$. 

{\bf Sketch of the proof of Proposition~\ref{p:w-v-n-dirac-dim-3}:} We choose a smooth function 
$\phi _{el} : ]0; +\infty[\dans\R$ tending to $0$ at infinity and having a finite limit at $0$. 
We choose smooth $\rL^2$-functions $u_1, u_2 : ]0; +\infty[\dans\R$ such that $u_1^2+u_2^2$ never 
vanishes, 
\begin{equation}\label{eq:cond-infini}
\lim _{r\to +\infty}\left\|\frac{d}{dr}\left(
\begin{array}{c}
u_1\\
u_2
\end{array}
\right)\right\|\cdot \left\|\left(
\begin{array}{c}
u_1\\
u_2
\end{array}
\right)\right\|^{-1}\ =\ 0\, ,
\end{equation}
for some norm $\|\cdot\|$ on $\R^2$, and the following $\rho _0$-dependent condition is satisfied. Recall that $\kappa =\kappa _{\rho _0}$ is an nonzero integer. 
If $\kappa >0$, we require that, near $0$, $u_1=0$ and $u_2=r^{\kappa}$. If $\kappa <0$, we impose that, near $0$, $u_1=r^{-\kappa}$ and $u_2=0$. \\
Note that functions $u_1$ and $u_2$ that satisfy, near infinity, $u_1=r^{-\delta}$ and $u_2=r^{-\delta}$ with $\delta >1/2$, also satisfy \eqref{eq:cond-infini} and are 
$\rL^2$ at infinity. \\
Let $u_1$ and $u_2$ satisfy the above conditions. We note that the derivatives $u_1'$ and $u_2'$ are also in $\rL^2$. Using the precise form of $u_1$ and $u_2$ near zero, 
we find that the two fractions in \eqref{eq:comp-V-1} are $o(r^0)$ and $\pm 1+o(r^0)$, respectively, as $r\to 0$. The two fractions in \eqref{eq:comp-V-2} are 
$\kappa r^{-1}+o(r^0)$ and $o(r^0)$, respectively, as $r\to 0$. Thus, the functions $\phi _{sc}$ and $\phi _{am}$ have finite limit at $0$. Since the matrix $\exp(rM)$ is 
invertible for any $r>0$ and $u_1^2+u_2^2$ does not vanish, $v_1^2+v_2^2$ never vanishes. Therefore $\phi _{sc}$ and $\phi _{am}$ are smooth everywhere. Since the functions 
$r\donne \exp(\pm rM)$ are bounded, we see that, thanks to \eqref{eq:cond-infini}, $\phi _{sc}$ and $\phi _{am}$ tend to $0$ at infinity. In particular, we see that  
$V(Q)(\D _0+i\un _4)^{-1}$ is compact. Now standard results show that 
$\D$ is self-adjoint on $\Hc^1$, the domain of $\D_0$ (cf. \cite{th}), and 
$\sigma _{\rm ess}(\D)=\sigma (\D_0)$. Finally, one can check that the bound state $f$ belongs to $\Hc^1$. \qed

Now we come back to the Schr\"odinger operator to mention a result by \cite{ruu}. For the one-dimensional Schr\"odinger operator 
on the half-line, with a Dirichlet boundary condition, the authors provide an explicit construction of a potential, that can be 
complex, to get a finite number $n$ of embedded eigenvalues at given values. The method is similar to those used in \cite{rs4} but treat 
the different eigenvalues together in a sort of parallel computation. The oscillating part of the potential is now a sum of  
$\sin (k_jx)$, each $k_j$ being chosen to get one eigenvalue at a given value. Furthermore, the potential depends on $n$ 
additional, quite arbitrary parameters. On one hand, these paramaters are used to avoid the above problem of possible singularities 
in the potential and, on the other hand, they provide several potentials that produce the same set of embedded eigenvalues. 

The previous result shows that, on the half-line, a finite set of embedded eigenvalues at precise places can be preserved when the potential is changed 
in an appropriate way. To end this subsection, we present another interesting result of ``stability''. It concerns an embedded eigenvalue for the 
one-dimensional Schr\"odinger operator on $\R$. It is due to \cite{chm}. 
\begin{theorem}\label{th:stabilite}\cite{chm}. 
Let $k>0$ and $\gamma\in\R$ such that $|\gamma|>k$. Then, there exists a codimension one submanifold $M$ of $\rL^1(\R; \R)$ such 
that, for all $V\in M$, the operator $H=P^2+\gamma x^{-1}\sin (kx)+V(x)$ is self-adjoint in $\rL^2(\R)$ and it has an  
eigenvalue at $k^2/4$ which is embbedded in the essential spectrum that is given by $[0; +\infty[$. 
\end{theorem}
The result in \cite{chm} actually holds true for non real potentials and also for $H$ viewed as an unbounded operator 
on $\rL^p(\R)$ with $1\leq p<\infty$. As we shall see below, the full result is in fact stronger, even in the 
$\rL^2(\R)$ self-adjoint setting.

\subsection{Dense set of embedded eigenvalues.} 
\label{ss:dense}

In the previous subsection, we saw that appropriate oscillating potentials can produce one embedded eigenvalue in the continuous 
spectrum. Here we present results showing that similar oscillating potentials can even produce a dense set of embedded eigenvalues. 
First results of this kind were apparently obtained by Naboko in \cite{n} for Schr\"odinger and Dirac operators in one dimension. 
While Naboko considered Dirac operators with a scalar potential, Schmidt provided a similar result for electrostatic potentials 
in \cite{sc}. All those results concern the line and the half-line with boundary conditions and need a condition of rational 
independence of the eigenvalues. We present below a result of this kind that does not require any rational 
independence. It is due to Simon (see \cite{si2}) and uses an appropriate serie of Wigner-Von Neumann potentials. 

\begin{theorem}\label{th:dense-point-spec}\cite{si2}. 
Let $(\kappa _n)_{n\in\N}$ be an arbitrary sequence of distinct positive numbers. Let $g : [0; +\infty[\dans\R$ be an increasing 
function such that $g(0)=1$ and $\displaystyle\lim_{+\infty}g=+\infty$. Then there exists an even, real potential $V$ on $\R$ such that, for each 
$n\in\N$, $\kappa _n$ is an eigenvalue of $P^2+V$ and, for all $x\in\R$, $|V(x)|\leq g(|x|)(1+|x|)^{-1}$. 
\end{theorem}
Simon provided also a version of this result for the half-line with boundary conditions (see \cite{si2}). We note that one can 
choose a function $g$ as above such that, for $a>3/4$, there exists $C>0$ such that, for all $x\in\R$, $|V(x)|\leq C(1+|x|)^{-a}$. 
In particular, a result by Kiselev (see \cite{ki}) shows in that case that $[0; +\infty[$ is the support of the absolutely 
continuous spectrum of $-\Delta _x+V$. Therefore, one can have in that case a dense point spectrum embedded in the absolutely 
continuous spectrum. 

For appropriates sequences $(R_n)_{n\in\N}$ and $(\varphi_n)_{n\in\N}$ with $\lim R_n=+\infty$, for some real valued 
potential $W$ supported in $[0; 1]$, the potential $V$ 
in Proposition~\ref{th:dense-point-spec} is, for $x\in\R$, given by the pointwise finite sum  
\[V(x)\ =\ W(x)\, +\, 4\sum _{n=0}^\infty\kappa_n\, \un_{\{\cdot >R_n\}}(x)\, 
\frac{\sin\bigl(2\kappa_nx+\varphi_n\bigr)}{x}\, ,\]
where $\un _A$ denotes the characteristic function of the set $A$. Note that the sinus functions $\sin 
(2\kappa_nx+\varphi_n)$ can be written as $n$-dependent linear combinaison of $\sin (2\kappa_nx)$ and 
$\cos (2\kappa_nx)$. Thus, each term in the above sum is, for large $x$, equal to a linear combination of 
a potential \eqref{eq:W} and of a cosinus version of a potential \eqref{eq:W}.

\subsection{Absence of embedded eigenvalues.} 
\label{ss:absence}

The results of the previous subsections could give the impression that it is easy to produce an embedded eigenvalue in the 
continuous spectrum by using an oscillating potential. We will see in this subsection that this is not true and one has to 
choose carefully the oscillating potential. 

Let us start with a result, due to Froese and Herbst, on the absence of positive eigenvalue for Schr\"odinger 
operators. 
\begin{theorem}\label{th:no-positive-eigenv}\cite{fh}. 
Consider a function $V : \R^d\dans\R$ such that $V(Q)(H_0+1)^{-1}$ and $(H_0+1)^{-1}Q\cdot\nabla V(Q)(H_0+1)^{-1}$ 
are compact and such that, for any $\epsilon>0$, there exists $C_\epsilon>0$ such that, as quadratic forms, 
\[Q\cdot\nabla V(Q)\ \leq \ \epsilon H_0\, +\, C_\epsilon\, .\]
Then $H=H_0+V(Q)$ has no positive eigenvalue. 
\end{theorem}
In fact, Froese and Herbst treat the more difficult case of $N$-body Schr\"odinger operators and also proved 
a slightly better result, even in the two-body case. 

We note that if $V=W_{\alpha\beta}+V_{lr}$ with $\beta>\alpha$ (see \eqref{eq:W}) and a long range potential $V_{lr}$, 
then the above result applies since $W_{\alpha\beta}$ is also a long range potential. In that case, the oscillating part 
$W_{\alpha\beta}$ of the potential does not produce any positive embbedded eigenvalue. 

To produce a positive embbedded eigenvalue, one needs to give a sufficient strength to the oscillating part. 
Indeed, it was proved in \cite{fh} (cf. Corollary 2.6) that, if $V=W_{11}+V_{lr}$ with $|w|<k$, then $H$ 
has no positive eigenvalue, in dimension $1$. A similar result is provided in \cite{chm}. In the 
framework of Theorem~\ref{th:stabilite}, $H$ has no positive eigenvalue if $|w|<k$. 

Furthermore, one has to choose carefully the oscillations to get a positive embbedded eigenvalue. Among the 
oscillating potentials of the form \eqref{eq:W}, it seems that one has to avoid the case $\alpha >1$, as suggested by the 
following result. 
\begin{theorem}\label{th:no-positive-eigenv-osci}\cite{jm}. 
Let $V : \R^d\dans\R$ be given by $V=W_{\alpha\beta}+V_{sr}$ with $\alpha >1$ and $\beta>1/2$. 
Then $H=H_0+V(Q)$ has no positive eigenvalue. 
\end{theorem}
This result is still valid if one adds to $V$ an appropriate long-range potential and $H_0$-compact, local singularities. 

An inspection of the proof of Theorem~\ref{th:no-positive-eigenv-osci} shows that a key argument is provided by the 
following compacity result, that does not hold true when $\alpha =1$. 
\begin{proposition}\label{p:oscillations-alpha>1} {\rm[}\cite{jm}, Proposition 2.4{\rm ]}. 
Let $\alpha >1$ and $d\geq 1$. Then, for any $p\geq 0$, there exist $\ell _1\geq 0$ and $\ell _2\geq 0$ such that 
$\langle P\rangle^{-\ell _1}\langle Q\rangle ^p(1-\kappa (|Q|))\sin (k|Q|^\alpha)\langle P\rangle^{-\ell _2}$ 
extends to a compact operator on $\rL^2(\R^d)$. 
\end{proposition}  
In the radial case, White even proved the absense of positive eigenvalue if $\alpha\neq 1$ (cf. \cite{w}).

We saw in Subsection~\ref{ss:existence} that one can use several oscillating potentials to get several embedded eigenvalues. 
Is that really necessary? For the potentials \eqref{eq:W}, it is essentially true. To explain what we mean, 
notice first that, in Subsection~\ref{ss:existence}, a $\rL^1$ potential was always added 
to the oscillating potential to produce an embedded eigenvalue. 
For such additional potentials, one can prove that the Wigner Von Neumann potential $W_{11}$ (cf. \eqref{eq:W}) 
produce at most one embedded eigenvalue (see \cite{fh,chm}).

\section{Limiting absorption principles.} 
\label{s:tal}
\setcounter{equation}{0}

This section is devoted to the second topic we want to consider, namely the limiting absorption principle (LAP)
for Schr\"odinger and Dirac operators. See \eqref{eq:tal-A}, \eqref{eq:tal-Q}, \eqref{eq:tal-A-dirac}, and 
\eqref{eq:tal-Q-dirac}. 

\subsection{The radial case.} 
\label{ss:ode}

In this subsection, we focus on LAPs for the Schr\"odinger operators obtained by technics of ordinary differential equations. 
They apply in one dimension and also for the radial, multidimensional case. 

In the papers \cite{bd,dmr,ret1,ret2}, the LAP \eqref{eq:tal-Q} was derived on compact intervals $\Ic$ 
included in $]0; +\infty[$ and avoiding a certain discrete set. The potential contains an oscillating part 
of the form \eqref{eq:W} for appropriate values of $(\alpha ; \beta )$ (see Figure~\ref{dessin:1} below) and 
a radial long range part. Using a pertubative argument, one can show that this result is preserved if one 
adds a (non necessary radial) short range potential. \\
The main drawback of this approach is that it cannot be applied when a non-radial long range potential is present. 
Despite this restriction, cases of the above result are not covered, up to now, by another method. 
This is the case when $\alpha =1$, $2/3<\beta <1$, and $\Ic$ is close and above $k^2/4$ (see Subsection~\ref{ss:putnam} 
below). Furthermore, technics of ordinary differential equations provide many interesting features on the 
oscillating potential (see also Appendix 2 to Section XI.8 in \cite{rs3} and also \cite{chm,w}). \\
We point out that a weak version of the LAP \eqref{eq:tal-Q} in the radial case is obtained in \cite{w} for a larger 
set of values of $(\alpha ; \beta )$.

\subsection{Local Putnam-Lavine theory.} 
\label{ss:putnam}

In this subsection, we prepare the treatment of the general case, i.e. when the potential is not radial, for the Schr\"odinger and  Dirac operators. Here we review several methods to get the LAP. Each method is actually applicable to treat some potentials \eqref{eq:W} but not all. 

Historically, LAPs for Schr\"odinger operators were first obtained by pertubation, starting from the LAP for 
the Laplacian $H_0$ (see \cite{rs4}, p. 172-177). Actually, this can be applied here when $V=W_{\alpha\beta}+V_{sr}$ with $\beta >1$ 
since, in this case, $W_{\alpha\beta}$ is of short range. This does not work however if $V$ contains a long range 
part. The reason is essentially the following. There exists $\epsilon>0$ such that 
$V_{sr}(Q)\langle Q\rangle^{\epsilon+1}$ is $H_0$-compact while for any $\epsilon>0$, 
$V_{lr}(Q)\langle Q\rangle^{\epsilon+1}$ is not $H_0$-compact. 
In \cite{co,cg}, the LAP was proved pertubatively for a class of oscillatory potentials that 
take the form $V(x)=x\cdot\nabla W(x)+V_{sr}(x)$ for short range $W$ and $V_{sr}$. This includes the 
case $\alpha+\beta >2$ in the present situation. \\
To go beyond, non pertubative methods had to be developped. 
Lavine initiated nonnegative commutator methods in \cite{l1,l2} by adapting Putnam's idea (see \cite{cfks} p. 60). 
Mourre introduced 1980 in \cite{m} a powerful, commutator method, nowadays called
``Mourre commutator theory'' (see \cite{abg,cfks,ggm,jmp,sa}). These methods have some idea in common 
(positivity of some commutator $i[H,B]$) but the latter is more flexibel and also provides stronger 
results on the behaviour of the resolvent of $H$ near the spectrum. They both apply here when 
$V=W_{\alpha\beta}+V_{sr}+V_{lr}$ with $\beta >\alpha$ since $W_{\alpha\beta}$ is actually a long range potential in that 
case. Refined versions of Mourre's commutator method can be used to cover the previous results (see \cite{jm,mar}).  
But it was proved in \cite{gj2,jm} that the required assumptions for these methods are not satisfied 
when $\alpha =1$ and $0<\beta<1$. Roughly speaking, the problem comes from the fact that the commutator $i[H,A]$ does not ``contain enough decay in $\langle Q\rangle$'' (see details below). We refer to the books \cite{abg,cfks} for details on Mourre's commutator method. We also mention an approach of the LAP via spectral measures in \cite{ben}, which seems however to have difficulties to treat a long range 
part in the potential. 

In Subsections~\ref{ss:almost-short-range} and~\ref{ss:weak-decaying} below, we shall make use of the ``local Putnam-Lavine theory'' to get the LAP for a larger set of parameters $(\alpha ; \beta)$, for Schr\"odinger and Dirac operators. We present here an abstract version of this theory. Since it is related to Mourre's commutator theory, we sketch the latter a little bit more. 

The ``local Putnam-Lavine theory'' was introduced in \cite{gj}. 
Then it was slightly modified and simplified by C. G\'erard in \cite{ge}. In \cite{gj2}, it was proved that G\'erard's version 
actually works under quite weak assumptions (in particular weaker than those in \cite{ge}) and was called ``Weighted 
Mourre theory''. This name is not appropriate since the corresponding theory does not make use of differential 
inequalities, which were a cornerstone in the usual Mourre commutator theory (see below). In fact, the approach in \cite{gj,ge,gj2} only 
use a kind of positivity of operators of the form $\theta (H)[H, iB]\theta (H)$, for some self-adjoint operators $B$ 
and for some localisation functions $\theta$. This is reminiscent of the works of Putnam (see \cite{cfks}, p. 61) 
and Lavine (see \cite{l1,l2}) but introduces localisations in $H$ as a new feature. It is thus natural to call it ``local Putnam-Lavine theory''. \\
Let us now review and compare Mourre's commuator theory and the ``local Putnam-Lavine theory''. \\
Let $T$ be a self-adjoint operator acting in some Hilbert space $\Hbb$ ($T$ will play 
the r\^ole of $H$ in our setting). Assume that there is another self-adjoint operator $S$ 
in $\Hbb$ such that, in some sense, the commutator $[T, S]$ is defined on large enough domain in $\Hbb$. Since $T$ and $S$ are a priori unbounded, this is not a trivial requirement. The notion of $C^1(S)$ regularity provides an appropriate framework for this situation (cf. \cite{abg}). 
Although this point is important, we shall not enter into details here. Assume further that, on some compact interval $\Jc$ of $\R$, we have the following Mourre estimate, in the 
form sense, 
\begin{equation}\label{eq:mourre-T-S}
E_\Jc(T)[T; iS]E_\Jc(T)\ \geq \ cE_\Jc(T)\, +\, K\, ,
\end{equation}
where $c>0$, $K$ is a compact operator on $\Hbb$, and $E_\Jc(T)$ is the spectral measure 
of $T$ on $\Jc$. If the commutator $[T; S]$ is ``nice enough'' (correctly, if $T\in C^1(S)$), then the Virial Theorem states that the point spectrum of $T$ in $\Jc$ is finite. 
This is the first step in Mourre's commutator theory. To get a LAP for $T$ of the 
form 
\begin{equation}\label{eq:tal-T-S}
\sup_{\Re z\in\Ic,\atop\Im z\neq 0}\bigl\|\langle S\rangle^{-s}(T-z)^{-1}\langle S\rangle^{-s}\bigr\|\ <\ +\infty\, ,
\end{equation}
for some $s>0$, on some $\Ic\subset\Jc$ ($\Ic$ necessarily avoids the point spectrum of $T$), one introduces modified resolvents $0<\epsilon\donne (T+T_\epsilon-z)^{-1}$, where $T_\epsilon$ ``tends'' to $0$ as $\epsilon\to 0$ and ``contains'' $[T; S]$. The LAP is 
derived from differential inequalities for the previous function of $\epsilon$ and this 
is the second step of Mourre's commutator theory. In fact, one gets a more precise result on the resolvent of $T$ near $\Ic$ than the LAP. This second step requires a better 
``regularity'' of $T$ w.r.t. $S$ than the one required by the Virial theorem. 
The latter regularity is granted when $T=H$, $S=A$, and $\alpha=\beta=1$, but not the former (cf. \cite{gj2}). \\
The ``local Putnam-Lavine theory'' aims to get \eqref{eq:tal-T-S} when less regularity 
than the one required by Mourre's commutator theory is available. Since one has to find an 
interval $\Ic$ that avoids the point spectrum of $T$, one also starts by proving 
a Mourre estimate \eqref{eq:mourre-T-S} to get the Virial Theorem on $\Jc$ and choses $\Ic$ inside $\Jc$. Then one applies the 
\begin{theorem}\label{th:LAP-T-S}\cite{gj2}. 
Under a suitable regularity assumption on $T$ w.r.t. $S$, the LAP \eqref{eq:tal-T-S} for $s>1/2$ on $\Ic$ is valid if there exist an interval $\Jc$, containing $\Ic$ in its interior, and a real valued, bounded function $\varphi$ such that the 
``weighted Mourre estimate'' for $T$ on $\Jc$, given by 
\begin{equation}\label{eq:weighted-mourre-T-S}
E_\Jc(T)[T; i\varphi (S)]E_\Jc(T)\ \geq \ E_\Jc(T)\langle S\rangle^{-2s}E_\Jc(T)\, ,
\end{equation}
holds true. 
\end{theorem}
In \cite{ge}, it was proven that \eqref{eq:weighted-mourre-T-S} follows from the Mourre estimate \eqref{eq:mourre-T-S}, if $T$ has a good enough regularity w.r.t. $S$ (actually a much better regularity than the one required by Mourre's commutator theory). This means, in particular that, if this regularity is good enough to apply Mourre's commutator theory, the ``local Putnam-Lavine theory'' becomes useless since Mourre's commutator theory will give better results. But, in 
concrete applications, one can hope to directly prove \eqref{eq:weighted-mourre-T-S} when 
the regularity required by Mourre's commutator theory is not available or difficult to check. This is precisely what happens when $T=H$, $S=A$, and $\alpha=\beta=1$ (see 
Subsection~\ref{ss:almost-short-range}). \\
However, as we shall see in Subsection~\ref{ss:weak-decaying}, one faces sometimes the following difficulty when one tries to show \eqref{eq:weighted-mourre-T-S}.  There is no spectral limitation on $S$ in \eqref{eq:weighted-mourre-T-S}. This estimate is local in $T$ but not in $S$. For instance, if one considers \eqref{eq:weighted-mourre-T-S} for $T=H_0$ and 
$S=Q$, in dimension $1$, the estimate is localised in the Fourier variable but not in the original variable. \\
To illustrate this difficulty, let us describe a similar problem in an simple framework. Consider a positive, continuous function $f$ on $\R$. We want to find 
a positive lower bound for $f$. This is not possible in general but, if we further know that the ``liminf'' of $f$ at $-\infty$ and $+\infty$ are positive, then $f$ indeed has a positive lower bound. Without further informations on $f$, we cannot guess such a bound but we can prove its existence by finding a positive lower bound for $f$ outside some compact set and then by arguing that a continuous, positive function on a compact set does have a positive lower bound. \\
To prove \eqref{eq:weighted-mourre-T-S} in applications, it might be difficult to guess an appropriate function $\varphi$. However, we can try the strategy described in the above, simple problem. This is exactly what the primitive version of the ``local Putnam-Lavine theory'' in \cite{gj} does. One tries to prove \eqref{eq:weighted-mourre-T-S} for ``large'' $S$ (see Proposition 3.1 in \cite{gj}) and to derive \eqref{eq:weighted-mourre-T-S} for ``bounded'' $S$ from the 
Mourre estimate \eqref{eq:mourre-T-S} (see Proposition 3.6 in \cite{gj}). \\
To fulfil the previous idea, it is convenient to reformulate the LAP \eqref{eq:tal-T-S} in the following way:\\
A weighted Weyl sequence associated to $T$ on $\Ic$ with weights $\langle S\rangle^{-s}$ is a sequence $(f_n, z_n)_{n\in\N}$ such that, for all $n$, $z_n\in\C$, $\Im z_n\neq 0$, $\Re z_n\in\Ic$,  $E_\Ic (T)f_n=f_n\in\Dc (T)$, and $(T-z_n)f_n\in\Dc(\langle S\rangle^s)$, and such that $\Im (z_n)\to 0$, $\|\langle S\rangle^s(T-z_n)f_n\|\to 0$, and $(\|\langle S\rangle^{-s}f_n\|)_{n\in\N}$ converges to some $\eta\geq 0$. Such $\eta$ is called the mass of the weighted Weyl sequence. Note that, if $s=0$, such a sequence $(f_n, z_n)_{n\in\N}$ is a Weyl sequence for $T$ in the usual sense. 
\begin{theorem}\label{th:LAP-T-S-special-sequences}\cite{gj2}. 
Under some (quite weak) regularity of $T$ w.r.t. $S$, the LAP \eqref{eq:tal-T-S} on $\Ic$ for $s\geq 0$ is equivalent to the property: for any weighted Weyl sequence $(f_n, z_n)_{n\in\N}$ associated to $T$ on $\Ic$ with weights $\langle S\rangle^{-s}$, the mass $\eta$ is zero.
\end{theorem}
Now, to prove the LAP \eqref{eq:tal-T-S} on $\Ic$ for $s>1/2$, one considers 
any weighted Weyl sequence $(f_n, z_n)_{n\in\N}$ associated to $T$ on $\Ic$ with weights $\langle S\rangle^{-s}$ and show that its mass $\eta$ is zero. 
At a technical level, one separates the ``large'' $S$ region from the ``bounded'' $S$ region by considering, for a compactly supported cut-off function $\chi$ that equals $1$ near $0$, the $\chi (S)f_n$, on one hand, and the 
$(1-\chi )(S)f_n$, on the other hand. More precisely, one uses a version of 
\eqref{eq:weighted-mourre-T-S} for ``large'' $S$ to prove that the mass 
of $((1-\chi )(S)f_n, z_n)_{n\in\N}$ is zero and the Mourre estimate \eqref{eq:mourre-T-S} to show that the mass of $(\chi (S)f_n, z_n)_{n\in\N}$ is also zero. This will be illustrated in details in Subsection~\ref{ss:weak-decaying}.

\subsection{Schr\"odinger and Dirac operators with an almost short range, oscillating potential.}
\label{ss:almost-short-range}

In this Subsection, we apply the ``local Putnam-Lavine Theory'' to get the LAP 
\eqref{eq:tal-A} for Schr\"odinger operators and the LAP \eqref{eq:tal-A-dirac} 
for Dirac operators, when the potential contains an almost short-range oscillating part. More precisely, this oscillating part will be given by the potential $W_{11}$  (see \eqref{eq:W} with $\alpha =\beta=1$). \\
We point out that, in this situation for Schr\"odinger operators, it was proved in \cite{gj2}, that the Hamiltonian $H$ does not have the regularity w.r.t. $A$, that is required by Mourre's commutator theory. We expect that this is also true for 
the corresponding Dirac operator. \\
Here we shall use the version of the ``local Putnam-Lavine Theory'' that is based on a weighted Mourre estimate and summarized in Theorem~\ref{th:LAP-T-S}. \\
Let us first consider the Schr\"odinger case with $V=W_{11}+V_{lr}+V_{sr}$. 
We only sketch the arguments to prove the LAP \eqref{eq:tal-A} on some compact interval $\Ic\subset ]0; +\infty[$. The details are provided in \cite{gj2} (see the proof of Theorem 4.15, there). \\
As mentioned above, we know that $\Ic$ should avoid the point spectrum of $H$. The latter could however be dense in 
the essential spectrum $]0; +\infty[$ (cf. Subsection~\ref{ss:dense}). As in Mourre's commutator theory, we first  
show a Mourre estimate on some appropriate interval $\Jc$ and then prove that the point spectrum of $H$ in $\Jc$ is finite. 
It is then possible to choose inside $\Jc$ a compact interval $\Ic$ that avoids the point spectrum. On such $\Ic$, we shall be able to prove a LAP. \\
In view of Subsection~\ref{ss:putnam}, it is natural to take for $S$ the operator 
$A=2^{-1}(P\cdot Q+Q\cdot P)$ and to look for a compact interval 
$\Jc$ (inside $]0; +\infty[$), for which we can prove a Mourre estimate \eqref{eq:mourre-T-S}. In the present framework, the Mourre estimate on $\Jc$ 
is valid if there exists $c>0$ and a compact operator $K$ (on $\rL^2(\R^d)$) such that 
\begin{equation}\label{eq:esti-mourre-J}
E_\Jc (H)[H, iA]E_\Jc (H)\ \geq\ c\, E_\Jc (H)\, +\, K\, ,
\end{equation}
where $\Omega\donne E_\Omega(H)$ is the spectral resolution of $H$. \\
It is well known that the quadratic form $[H_0+V_{lr}+V_{sr}, iA]$ extends to a bounded form on the domain 
$\Hc^2(\R^d)$ of $H_0$ and that \eqref{eq:esti-mourre-J} holds true if $[H, iA]$ is replaced by $[H_0+V_{lr}+V_{sr}, iA]$. 
Thus, we have to control the form $[W_{11}, iA]$ that is associated to the operator $-Q\cdot\nabla W_{11}(Q)$. 
This form does extend to a bounded form on $\Hc^2(\R^d)$ but not to a compact one, due to the fact that the 
function $x\donne x\cdot\nabla W_{11}(x)$ does not tend to $0$ at infinity. \\
To overcome this difficulty, we note that we may replace $E_\Jc (H)$ in \eqref{eq:esti-mourre-J} by $\theta (H)$ 
for some smooth, localised function $\theta$. Thus we have to control $\theta (H)[W_{11}, iA]\theta (H)$. 
Furthermore, since the difference $H-H_0$ is $H_0$-compact, it actually suffices to show that 
$\theta (H_0)[W_{11}, iA]\theta (H_0)$ is a compact form on $\rL^2(\R^d)$. We pick from \cite{gj2} the 
\begin{proposition}\label{p:oscillations-energy1}{\rm [}\cite{gj2}, Lemma 4.3, Lemma C.1, and Proposition A.1{\rm ]}. 
Take some $\theta\in \Cc_c^\infty (\R;\C)$ with small enough support in the interval $]0; k^2/4[$. Then, for any $\epsilon\in [0; 1[$, 
the operator $\theta (H_0)\langle Q\rangle ^\epsilon\sin (k|Q|)\theta (H_0)$ extends to a compact
operator on $\rL^2(\R^d)$. 
\end{proposition}  
\begin{rem}\label{r:oscillations-energie1}
In dimension $d=1$, Proposition~\ref{p:oscillations-energy1} still holds true if $]0; k^2/4[$ is replaced by 
$\R\setminus\{k^2/4\}$. It is more or less contained in \cite{fh}.  \\
In general, it is convenient to use an appropriate pseudodifferential calculus to prove 
Proposition~\ref{p:oscillations-energy1}. In this way, one can see that, up to a compact correction, the 
operator $\theta (H_0)\langle Q\rangle ^\epsilon e^{ik|Q|}\theta (H_0)$ is given by the Weyl quantization of 
the function $(x; \xi )\donne \langle x\rangle^{\epsilon}\theta (|\xi|^2)\theta (|\xi -k|^2)$. If the support of $\theta$ 
is small enough in $]0; k^2/4[$ (in $\R\setminus\{k^2/4\}$, if $d=1$), this function vanishes identically. 
If $d>1$ and the support of $\theta$ is contained in $[k^2/4; +\infty[$, this is not true anymore and the considered 
operator is not compact. \\
Proposition~\ref{p:oscillations-energy1} should be compared to Proposition~\ref{p:oscillations-alpha>1} to enlighten 
the difference between the cases $\alpha =1$ and $\alpha >1$. 
\end{rem}
Thanks to Proposition~\ref{p:oscillations-energy1}, we get the Mourre estimate \eqref{eq:esti-mourre-J} on any {\it small enough} compact interval $\Jc\subset ]0; k^2/4[$ ($\Jc\subset\R\setminus\{k^2/4\}$, if $d=1$). Using a compacity argument, we recover \eqref{eq:esti-mourre-J} on any compact interval $\Jc\subset ]0; k^2/4[$ ($\Jc\subset\R\setminus\{k^2/4\}$, if $d=1$). Then we can 
use the Virial Theorem (see \cite{abg}, p. 295) to show the finitness of the point spectrum of $H$ in $\Jc$. \\
Now we take a compact interval $\Ic$ inside $]0; k^2/4[$ ($\R\setminus\{k^2/4\}$, if $d=1$), that avoids the 
point spectrum of $H$. Since, for any $\delta>0$, $\Ic$ can be covered by a finite number of compact intervals 
of size $\delta$, we can deduce the LAP \eqref{eq:tal-A} on $\Ic$ from the LAP \eqref{eq:tal-A} on those 
intervals. In other words, we may assume $\Ic$ as small as we want. \\
To prove the LAP on such $\Ic$, it suffices, by Theorem~\ref{th:LAP-T-S}, 
to prove a weighted Mourre estimate of the type \eqref{eq:weighted-mourre-T-S}. 
We shall show that, for all 
$s>1/2$, there exist an intervall $\Jc$, that contains $\Ic$ in its interior and avoids the point spectrum of $H$, and a bounded, real function $\varphi$ on $\R$ such that
\begin{equation}\label{eq:esti-mourre-poids}
E_\Jc (H)[H, i\varphi (A)]E_\Jc (H)\ \geq\ E_\Jc (H)\langle A\rangle^{-2s}E_\Jc (H)\, .
\end{equation}
It is convenient to look for a function $\varphi$ of the form $\psi (\cdot /R)$, 
for some $R\geq 1$ and an appropriate real valued, bounded function $\psi$, and to show 
\begin{equation}\label{eq:esti-mourre-poids-R}
E_\Jc (H)[H, i\psi(A/R)]E_\Jc (H)\ \geq\ E_\Jc (H)\langle A/R\rangle^{-2s}E_\Jc (H)\, .
\end{equation}
Indeed, \eqref{eq:esti-mourre-poids} follows from \eqref{eq:esti-mourre-poids-R} since $\langle A/R\rangle^{-2s}\geq \langle A\rangle^{-2s}$. \\
To guess an appropriate function $\psi$, it is natural to express $\psi(A/R)$ with the help of Helffer-Sj\"ostrand formula 
\begin{equation}\label{eq:hs}
\psi(A/R)\ =\ \, \int_{\C}\, \partial _{\bar{z}}\psi^{\C}(z)(z-A/R)^{-1}\, d\mu (z)\, ,
\end{equation}
where $\mu$ is the Lebesgue measure on $\C$ and $\psi^{\C}$ is an almost-analytic extension of $\psi$ (see \cite{gj,gj2,hs} for details). 
Commutator expansions (see \cite{gj}) indicate that the commutator $[H, i\psi(A/R)]$ is essentially given by $\psi '(A/R)[H, iA/R]$ and 
one expects that the commutator $[H, iA]$ exhibits some positivity. Therefore, in view of the r.h.s. of \eqref{eq:esti-mourre-poids-R}, it 
is natural to choose $\psi$ such that $\psi'(t)=R\langle t\rangle^{-2s}$. Since $2s>1$, we may take 
\[\psi(t)\ =\ c\cdot R\cdot\, \int_{-\infty}^t\langle\tau\rangle^{-2s}\, d\tau\, ,\]
for some constant $c>0$ to be chosen later. 
However, the mentioned commutator expansions actually require at least an appropriate boundedness of the commutators 
$[H, A]$ and $[[H, A], A]$, that we do not have in the present setting. What we can use is the consequence of \eqref{eq:hs} 
given by 
\begin{equation}\label{eq:hs-commutateur}
[H, i\psi(A/R)]\ =\ \int_{\C}\, \partial _{\bar{z}}\psi^{\C}(z)\, (z-A/R)^{-1}[H, iA/R](z-A/R)^{-1}\, d\mu (z)\, .
\end{equation}
Now the first step is to replace each projection $E_\Jc (H)$ in the l.h.s. of \eqref{eq:esti-mourre-poids-R} by $\theta (H)$, for smooth localisation 
functions $\theta$, and to move a copy $\chi (H)$ of these localisations (with $\theta =\theta\chi$) inside the integral and through 
the resolvent $(z-A/R)^{-1}$. Then, one replaces these copies by localisations in $H_0$. Up to error terms, the l.h.s. of 
\eqref{eq:esti-mourre-poids-R} reduces to 
\[\theta (H)\Bigl(\int_{\C}\, \partial _{\bar{z}}\psi^{\C}(z)\, (z-A/R)^{-1}\chi (H_0)[H, iA/R]\chi (H_0)(z-A/R)^{-1}\, d\mu (z)\Bigr)
\theta (H)\, .\]
As in the proof of the Mourre estimate \eqref{eq:esti-mourre-J}, $\chi (H_0)[H, iA/R]\chi (H_0)$ is given by 
$R^{-1}\chi ^2(H_0)2H_0$ plus some ``small'' error term. Here, we used again Proposition~\ref{p:oscillations-energy1} to get rid of the 
contribution of the oscillating part $W_{11}$ of the potential. In the contribution of $R^{-1}\chi ^2(H_0)2H_0$ to the above integral, namely 
\[\theta (H)\Bigl(\int_{\C}\, \partial _{\bar{z}}\psi^{\C}(z)\, (z-A/R)^{-1}R^{-1}\chi ^2(H_0)2H_0(z-A/R)^{-1}\, d\mu (z)\Bigr)
\theta (H)\, ,\]
we move each localisation $\chi (H_0)$ through a resolvent $(z-A/R)^{-1}$ to get, up to some error term, 
\[\theta (H)\chi (H_0)\Bigl(\int_{\C}\, \partial _{\bar{z}}\psi^{\C}(z)\, (z-A/R)^{-1}R^{-1}[H_0, iA](z-A/R)^{-1}\, d\mu (z)\Bigr) 
\chi (H_0)\theta (H)\, ,\]
which is equal to $\theta (H)\chi (H_0)[H_0, i\psi (A/R)]\chi (H_0)\theta (H)$. Since the forms $[H_0, A]$ and $[[H_0, A], A]$ have 
nice boundedness properties, we can use the commutators expansions of \cite{gj} to write the above term, up to some error, as 
\begin{eqnarray*}
&&\theta (H)\chi (H_0)(\psi '(A/R))^{1/2}[H_0, iA/R](\psi '(A/R))^{1/2}\chi (H_0)\\
&=&c\, \theta (H)\chi (H_0)\langle A/R\rangle^{-s}
[H_0, iA]\langle A/R\rangle^{-s}\chi (H_0)\theta (H)\, .
\end{eqnarray*}
Now, we move again each $\chi (H_0)$ through a weight $\langle A/R\rangle^{-s}$ to transform this term into, up to some error term, 
\begin{eqnarray*}
&&c\, \theta (H)\langle A/R\rangle^{-s}
\chi (H_0)[H_0, iA]\chi (H_0)\langle A/R\rangle^{-s}\theta (H)\\
&=&c\, \theta (H)\langle A/R\rangle^{-s}\chi ^2(H_0)2H_0\langle A/R\rangle^{-s}\theta (H)\, ,
\end{eqnarray*}
which is bounded below by $cc'\theta (H)\langle A/R\rangle^{-s}\chi ^2(H_0)\langle A/R\rangle^{-s}\theta (H)$, for 
some $\Ic$-dependent constant $c'>0$. Moving again each localisation $\chi (H_0)$ through a weight $\langle A/R\rangle^{-s}$ and replacing each 
$\chi (H_0)$ by $\chi (H)$, this lower bound can be replaced by $cc'\theta (H)\langle A/R\rangle^{-2s}\theta (H)$, up to another error term, 
since $\theta =\theta\chi$. Choosing $c$ such that $cc'=2$, we arrive at 
\begin{equation}\label{eq:lower-bound}
\theta (H)[H, i\psi(A/R)]\theta (H)\ \geq\ 2\, \theta (H)\langle A/R\rangle^{-s}
\bigl(2{\rm I}\, +\, B\, \, +\, R^{-1}C\bigr)\langle A/R\rangle^{-s}\theta (H)\, ,
\end{equation}
where ${\rm I}$ is the identity operator on $\rL^2(\R^d)$ and where we put all the previous error terms in $B+R^{-1}C$. It turns out that the following properties are true. 
\begin{itemize}
 \item[a).] The operator $C$ is bounded and, although it depends on $\chi$ and $R$, its operator norm can be bounded above by some $\chi$-independent, $R$-independent constant.
 \item[b).]  The operator $B$ is a finite sum of terms of the form 
\[B_1\cdot K\cdot \chi (H_\sigma)\cdot B_2\, .\]
 \item[c).] $K$ is a compact operator on 
$\rL^2(\R^d)$ that depends neither on $\theta$, nor on $R$, but on $\chi$. 
 \item[d).] The operators $B_1$ and $B_2$ are bounded on $\rL^2(\R^d)$. 
 Although they do depends on $\chi$ and $R$, their operator norms can bounded above by some $\chi$-independent, $R$-independent constant.
\end{itemize}
Now we shrink the support of $\chi$ (and thus the one of $\theta$) to make the norm of $K\chi (H_\sigma)$ small enough. 
This is possible since we only have continuous spectrum of $H$ and $H_0$ in $\Jc$. Then we 
take $R$ large enough to ensure that the r.h.s 
of \eqref{eq:lower-bound} is bounded below by $\theta (H)\langle A/R\rangle^{-2s}\theta (H)$, yielding the weighted Mourre estimate 
\eqref{eq:esti-mourre-poids-R} on any small enouth $\Jc$, on which $\theta =1$. \\
We claimed that the contribution of the potential $V$ to the l.h.s. of \eqref{eq:lower-bound} can be bounded below by terms in 
$B+R^{-1}C$. This is quite straightforward (and standard) for the contributions of $V_{lr}$ and $V_{sr}$ since 
\[\langle Q\rangle^{\rho _{lr}/2}[V_{lr}(Q), iA]\langle Q\rangle^{\rho _{lr}/2}\ =\ -\langle Q\rangle^{\rho _{lr}/2}Q\cdot\nabla V_{lr}(Q)
\langle Q\rangle^{\rho _{lr}/2}\]
\[\mbox{and}\hspace{.4cm}\langle P\rangle ^{-1}\langle Q\rangle ^{\rho _{sr}/2}(V_{sr}(Q)A-AV_{sr}(Q))
\langle Q\rangle ^{\rho _{sr}/2}\langle P\rangle ^{-1}\]
are bounded. For the one of $W_{11}$, we saw that we essentially have to control 
\[\chi (H_0)[W_{11}(Q), A]\chi (H_0)\ =\ \chi (H_0)W_{11}(Q)A\chi (H_0)\, -\, \chi (H_0)AW_{11}(Q)\chi (H_0)\, .\]
This form extends to a compact form on $\rL^2(\R^d)$ thanks to Proposition~\ref{p:oscillations-energy1}. \\
Since \eqref{eq:esti-mourre-poids-R} implies 
\eqref{eq:esti-mourre-poids}, we can apply Theorem~\ref{th:LAP-T-S} to obtain the 
\begin{theorem}\label{th:tal-A-Wigner-gj2}\cite{gj2}. 
Take $V=W_{11}+V_{lr}+V_{sr}$ and a compact interval $\Ic\subset ]0;k^2/4[$ ($\Ic\subset\R\setminus\{k^2/4\}$, if 
$d=1$). Then, for any $s>1/2$, \eqref{eq:tal-A} holds true. 
\end{theorem}
\begin{rem}\label{r:tal-A-Wigner}
Actually, one can include in $V$ some local, $H_0$-compact singularities. 
\end{rem}
The above method can be recycled to treat the massive Dirac operator $\D$ with an oscillating part of type $W_{11}\un _4$ in the potential. 
To take into account the matrix structure of the Dirac operator, a natural candidate for the ``conjugate operator'' $A'$ is a bit 
more complicated than the operator $A$ we used for the Schr\"odinger case. To get the LAP \eqref{eq:tal-A} 
with $H$ replaced by $\D$ and $A$ replaced by $A'$ on some compact interval $\Ic$, it is convenient to choose a $\Ic$-dependent 
operator $A'$. Denote by $\Mcc _4(\C)$ the vector space of $4\times 4$-matrices with complex entries. Recall that $m>0$ denotes the mass. 
Let $\tau\in\Cc_c^\infty(\R; \R^+)$ such that $\tau =1$ on $\Ic$ and $\tau =0$ outside a slightly larger interval included in 
\[\bigl]-(m^2+k^2/4)^{1/2}; -m\bigr[\cup \bigl]m;(m^2+k^2/4)^{1/2}\bigr[\, .\]
We introduce the functions $\mu : \R^3\dans\R^+$, $F_j : \R^3\dans \R$, for $j\in\{1; 2; 3\}$, and 
$\Pi _\pm : \R^3\dans \Mcc _4(\C)$, defined by 
\[\mu (\xi)\ =\ \bigl(|\xi|^2+m^2\bigr)^{1/2}\, ,\hspace{.4cm}F_j (\xi)\ =\ \mu (\xi)^2\cdot |\xi|^{-2}\cdot \tau \bigl(\mu (\xi)\bigr)
\cdot\xi _j\]
\[\mbox{and}\hspace{.4cm}\Pi _\pm (\xi)\ =\ 2^{-1}\bigl(\un _4\, \pm\, \mu (\xi)^{-1}(\underline{\alpha}\cdot\xi\, +\, m\beta)\bigr)\, .\]
After conjugaison with the Fourier transform, the free Dirac operator \eqref{eq:free-dirac} acts as the multiplication by the 
matrix $\underline{\alpha}\cdot\xi\, +\, m\beta$ (where $\xi$ is the Fourier variable). The eigenvalues of this matrix are 
$\pm\mu (\xi)$, both with multiplicity two, and the corresponding spectral projections are $\Pi _\pm (\xi)$. \\
Let $F=(F_1; F_2; F_3)^T$. Following \cite{bmp}, we set 
\[\tilde{A}\ =\ 2^{-1}\bigl(Q\cdot F(P)\, +\, F(P)\cdot Q\bigr)\ =\ 2^{-1}\sum _{j\in\{1; 2; 3\}}\bigl(Q_j\cdot F_j(P)\, +\, F_j(P)\cdot Q_j\bigr)
\, ,\]
One can check that $[\mu (P), i\tilde{A}]=\mu (P)\tau (\mu (P))$. This fact should be compared to $[H_0; iA]=2H_0$. But one also needs to 
take into account the matrix structure of $\D_0$. It is thus natural to choose $A'$ as the self-adjoint realization in $\rL^2(\R^3; \C^4)$ of 
$\Pi _+(P)\tilde{A}\Pi _+(P)+\Pi _-(P)\tilde{A}\Pi _-(P)$. \\
Following the method we used to prove Theorem~\ref{th:tal-A-Wigner-gj2}, one can also prove the following result on the Dirac operator $\D$. 
\begin{theorem}\label{th:tal-A-Wigner-Dirac}\cite{mb}. 
Let $V=(W_{11}+V_{lr})\un _4+V_{sr}$, where the function $V_{lr}$ (resp. $V_{sr}$), defined on $\R^3$ with values in $\R$ (resp. the space of self-adjoint $4\times 4$-matrices), satisfies the long (resp. short) range assumption. 
Take a compact interval 
\[\Ic\, \subset\,  \bigl]m;(m^2+k^2/4)^{1/2}\bigr[\hspace{.4cm}\mbox{or}\hspace{.4cm}\Ic\, \subset\,  \bigl]-(m^2+k^2/4)^{1/2}; -m\bigr[\, .\]
Then, for any $s>1/2$, \eqref{eq:tal-A-dirac} holds true. 
\end{theorem}
%

\subsection{Schr\"odinger operators with a weak decaying, oscillating potential.}
\label{ss:weak-decaying}

Now, we come back to the Schr\"odinger operator and study the LAP for a another range of the parameters $(\alpha ; \beta)$ in the 
oscillating part \eqref{eq:W} of the potential. We shall focus on the region where $1\leq\alpha$, $1/2<\beta\leq 1$, and 
$\alpha +\beta\leq 2$. \\
Let us have a look at the method we followed to get Theorem~\ref{th:tal-A-Wigner-gj2}. It used in an important way the fact that the 
form $[W_{11}, iA]$ extends to a bounded form on $\rL^2(\R^d)$. Indeed, we used this property to prove the Mourre estimate 
\eqref{eq:esti-mourre-J}, to write down \eqref{eq:hs-commutateur}, and also each time we moved a localisation $\chi (H)$ through the 
resolvent $(z-A/R)^{-1}$. In the considered region for $(\alpha ; \beta)$, one can prove (cf. \cite{jm}) that the form $[W_{\alpha\beta}, iA]$ does not extend 
to a bounded form on $\Hc^2(\R^d)$. Therefore, we cannot simply follow the method. While we shall be able to recover the Mourre 
estimate, it seems difficult (at least complicated) to adapt the above arguments to get the weighted Mourre estimate \eqref{eq:esti-mourre-poids} 
to the present situation. \\
However, as we shall see now, it is possible to use the version of the ``local Putnam-Lavine Theory'', that was presented in \cite{gj} and sketched at the end of Subsection~\ref{ss:putnam} (see Theorem~\ref{th:LAP-T-S-special-sequences} and the comments following it). \\
For simplicity, we assume that $V=W_{\alpha\beta} +V_{sr}$ and sketch the arguments used in \cite{jm} to get the LAP.\\
We first need to show the Mourre estimate \eqref{eq:esti-mourre-J}. 
Consider again the above proof of 
\eqref{eq:esti-mourre-J}. We use the fact that $\beta >1/2$ to first show that 
\[\bigl(\theta (H)-\theta (H_0)\bigr)W_{\alpha\beta}A\theta (H_\sigma)\hspace{.4cm}\mbox{and}\hspace{.4cm}
\bigl(\theta (H)-\theta (H_0)\bigr)AW_{\alpha\beta}\theta (H_\sigma)\, ,\]
with $H_\sigma =H_0$ or $H_\sigma =H$, extend to compact forms on $\rL^2(\R^d)$. Thus, there exists a compact operator $K_0$ such that 
$\theta (H)[W_{\alpha\beta}, iA]\theta (H)=\theta (H_0)[W_{\alpha\beta}, iA]\theta (H_0)+K_0$. Now we use $\beta >0$ and 
Proposition~\ref{p:oscillations-alpha>1}, if $\alpha>1$, or Proposition~\ref{p:oscillations-energy1}, if 
$\alpha=1$, to see that $\theta (H_0)[W_{\alpha\beta}, iA]\theta (H_0)$ extends to a 
compact form on $\rL^2(\R^d)$. Thus we are able to perform the previous proof of the Mourre estimate 
\eqref{eq:esti-mourre-J}. We point out for latter purposes that, as usual, shrinking the size of $\Jc$, one 
deduces from it the following strict Mourre estimate: there exists $c>0$ such that 
\begin{equation}\label{eq:esti-mourre-strict-J}
E_\Jc (H)[H, iA]E_\Jc (H)\ \geq\ c\, E_\Jc (H)\, .
\end{equation}
Although the regularity assumptions needed by the Virial Theorem 
are not met (if $\beta<1$), its result is valid (see \cite{jm}). The main reason for this is the fact that a possible eigenvector of $H$ associated to an eigenvalue in $\Jc$ is nice enough to belong to the domain of $A$. Thus, 
the point spectrum of $H$ is finite in an interval $\Jc$ for which the Mourre 
estimate \eqref{eq:esti-mourre-J} is valid and it is actually empty on those 
intervals $\Jc$ for which the strict Mourre estimate \eqref{eq:esti-mourre-strict-J} holds true. \\
Instead of proving the LAP \eqref{eq:tal-A}, we shall prove the LAP \eqref{eq:tal-Q}. Note that if the latter is true for some $s>1/2$, 
it is also true for any $s'\geq s$, since $\langle Q\rangle^{s-s'}$ is bounded. Thus, we only need to show \eqref{eq:tal-Q} for 
$s>1/2$ and $s$ as close to $1/2$ as we want. Let $\delta =2s-1>0$, small enough. \\
Instead of proving the estimate \eqref{eq:esti-mourre-poids-R}, that is global in the position operator $Q$, we prove a similar estimate but ``at infinity in the position operator $Q$''. Let us make this precise. Let $\Jc$ be a compact interval 
inside $]0; k^2/4[$ ($\R\setminus\{k^2/4\}$ if $d=1$). Inside the interior of $\Jc$, take a compact interval $\Ic$ that avoids the 
(finite) point spectrum of $H$ in $\Jc$. Let $\Ec$ be a set of functions $f\in\rL^2(\R^d)$ satisfying $E_\Ic(H)f=f$ such that the 
function $\Ec\ni f\donne \|\langle Q\rangle^{-s}f\|$ is bounded. Let $\gamma =\beta -\delta>1/2$. \\
We claim that there exist $c_1>0$ and $R_1\geq 1$ such that, for all $R\geq R_1$, there exists a $H$-bounded, self-adjoint operator $B_R$ such 
that, for all $f\in\Ec$, 
\begin{eqnarray}
\label{eq:mourre-infini}
\langle f\, ,\, \bigl[H, iB_R\bigr]f\rangle&\geq& c_1\cdot \|\chi _R(Q)\langle Q\rangle^{-s}f\|^2\\
&&\, -\, O(R^{-\gamma})\cdot 
\|\chi _R(Q)\langle Q\rangle^{-s}f\|\, -\, O(R^{-\gamma -1})\, ,\nonumber
\end{eqnarray}
where $\chi _R(Q)=\chi (R^{-1}|Q|)$, $\chi\in\Cc^\infty(\R^+; \R)$, $\chi =0$ on $[0; 1]$, and $\chi =1$ on $[1; +\infty[$, and 
where the ``$O$'' are uniform w.r.t. the set $\Ec$. In fact, we can precisely take the operator $B_R=\chi _R(Q)^2g_{\delta}(Q)Q\cdot P+P\cdot Q
g_{\delta}(Q)\chi _R(Q)^2$, 
where 
\[g_{\delta}(x)\ =\ \bigl(2\, -\, \langle x\rangle^{-\delta}\bigr)\langle x\rangle^{-1}\, \geq \, 
\langle x\rangle^{-1}\, .\]
\eqref{eq:mourre-infini} plays the r\^ole of what we called the estimate 
\eqref{eq:esti-mourre-poids-R} for ``large'' $S$ with $S=Q$ in Subsection~\ref{ss:putnam}. The cut-off function $\chi _R(Q)$ indeed localises in a region 
where $|Q|$ is large. In contrast to \eqref{eq:esti-mourre-poids-R}, we allow 
error terms in \eqref{eq:mourre-infini}, that are however small for $R$ large. \\
To prove \eqref{eq:mourre-infini}, we write $f=\theta (H)f$, for an appropriate, smooth cut-off function $\theta$. We use the fact that $H$ and $H_0$ have a good regularity w.r.t. $Q$ in the sense that the forms $[H, Q_j]=[H_0, Q_j]$ and $[[H, Q_j], Q_k]=[[H_0, Q_j], Q_k]$ are $H$-bounded and bounded, respectively. Now \eqref{eq:mourre-infini} follows from the following facts: 
\begin{itemize}
 \item[a).] The contribution of $V_{sr}$ to \eqref{eq:mourre-infini} can be hidden in the two last terms, the errors terms. 
 \item[b).] Thanks to the form of $B_R$, 
\[[H_0, iB_R]\ =\ \chi _R(Q)g_\delta(Q)^{1/2}[H_0, iA]g_\delta(Q)^{1/2}\chi _R(Q)\, +\ \mbox{a nice error term}\, .\]
 \item[c).] Since $E_\Ic(H)f=f$, we may write $f=\theta (H)f$, for an appropriate, smooth cut-off function $\theta$. On the l.h.s. of \eqref{eq:mourre-infini}, we insert those functions $\theta (H)$ and move them from the exterior to the 
 centre. This creates terms that contribute to the error terms 
 on the r.h.s. of \eqref{eq:mourre-infini}. We replace each moved function $\theta (H)$ by $\theta (H_0)$ producing this way new terms that can be seen as error terms. 
 \item[d).] The main term on the l.h.s. of \eqref{eq:mourre-infini} is of the 
 form 
\[\langle f\, ,\, \chi _R(Q)g_\delta(Q)^{1/2}\theta (H_0)[H_0, iA]\theta (H_0)g_\delta(Q)^{1/2}\chi _R(Q)f\rangle\]
and can be bounded below by the first term on the r.h.s. of \eqref{eq:mourre-infini}. 
 \item[e).] It remains to explain why the contribution of $W_{\alpha\beta}$ to 
 the l.h.s. of \eqref{eq:mourre-infini} produces terms that belong to the 
 errors terms on the r.h.s. of \eqref{eq:mourre-infini}. This follows 
 from Proposition~\ref{p:oscillations-alpha>1}, if $\alpha>1$, and from Proposition~\ref{p:oscillations-energy1}, if $\alpha=1$. 
\end{itemize}
Now we follow the arguments in \cite{gj}. We shall apply Theorem~\ref{th:LAP-T-S-special-sequences} with $S=Q$ to get the LAP \eqref{eq:tal-Q}. Take a weighted Weyl sequence $(f_n, z_n)_{n\in\N}$ associated to $H$ on $\Ic$. Recall that, for all $n$, $z_n\in\C$, $\Im z_n\neq 0$, $\Re z_n\in\Ic$,  $E_\Ic (H)f_n=f_n\in\Dc (H)$, and $(H-z_n)f_n\in\Dc(\langle Q\rangle^s)$, and such that $\Im (z_n)\to 0$, 
$\|\langle Q\rangle^s(H-z_n)f_n\|\to 0$, and $(\|\langle Q\rangle^{-s}f_n\|)_{n\in\N}$ converges to some $\eta\geq 0$, the mass of the weighted Weyl sequence. 
We show that $\eta =0$. \\
For all $n$, $\theta (H)f_n=f_n$. For $\Ec=\{f_n; n\in\N\}$, we 
have \eqref{eq:mourre-infini}. Using the properties 
\begin{equation}\label{eq:prop-f_n}
\|\langle Q\rangle^s(H-z_n)f_n\|\to 0\hspace{.4cm}\mbox{and}\hspace{.4cm}
\|\langle Q\rangle^{-s}f_n\|\to\eta\, ,
\end{equation}
and expanding the commutator, we can show that the l.h.s. 
of \eqref{eq:mourre-infini} tends to zero as $n\to\infty$. This implies that, for $R\geq R_1$,  
\begin{equation}\label{eq:control-infini-f_n}
\limsup _{n\to\infty}\|\chi_R(Q)\langle Q\rangle^{-s}f_n\|\ =\ O(R^{-\gamma})\, .
\end{equation}
Let $\tau_R=1-\chi_R$. We apply the strict Mourre estimate \eqref{eq:esti-mourre-strict-J} to each 
$\tau_R(Q)f_n$. Thanks to the good boundedness properties of the commutators $[H, Q_j]=[H_0, Q_j]$ and 
$[[H, Q_j], Q_k]=[[H_0, Q_j], Q_k]$, we can move the $\tau_R(Q)$ into the commutator and 
control the error terms with the help of \eqref{eq:control-infini-f_n}. This leads to 
\begin{equation}\label{eq:control-C_R-f_n}
\langle f_n\, ,\, [H, iC_R]f_n\rangle\ \geq \ c\|\tau_R(Q)f_n\|^2\, -\, O(R^{s-\gamma})\|\tau_R(Q)f_n\|\, -\, 
O(R^{2s-2})\, ,
\end{equation}

for some $R$-dependent bounded operator $C_R$. Expanding the commutator on the l.h.s. of \eqref{eq:control-C_R-f_n} and using again \eqref{eq:prop-f_n}, we prove that this l.h.s. tends to zero as $n\to\infty$. This yields 
\[\limsup _{n\to\infty}\|\tau_R(Q)f_n\|\ =\ O(R^{s-\gamma})\, ,\]
with $s-\gamma<0$. Combining this with \eqref{eq:control-infini-f_n}, we obtain that 
\[\eta \ =\ \lim _{n\to\infty}\|\langle Q\rangle^{-s}f_n\|\ =\ 0\, .\]
By Theorem~\ref{th:LAP-T-S-special-sequences}, this proves the LAP \eqref{eq:tal-Q} on $\Ic$. We just have sketched the proof of the following 
\begin{theorem}\label{th:tal-Q-jm}\cite{jm}.
Let $1/2<\beta \leq 1\leq\alpha$ and $V=W_{\alpha\beta}+V_{sr}$. Take a compact interval 
$\Ic\subset ]0;k^2/4[$ ($\Ic\subset\R\setminus\{k^2/4\}$, if 
$d=1$). Then, for any $s>1/2$, \eqref{eq:tal-Q} holds true. 
\end{theorem}
\begin{rem}\label{r:tal-Q-jm}
Theorem~\ref{th:tal-Q-jm} remains valid if $V$ contains $H_0$-compact, local singularities and a long range part 
$V_{lr}$, provided that $\beta +\rho _{lr}>1$ (see \cite{jm}). However, we believe that the condition 
$\rho _{lr}>0$ should be sufficient.\\
In technical terms, the main difference between the proof of Theorem~\ref{th:tal-A-Wigner-gj2} and the one 
of Theorem~\ref{th:tal-Q-jm} takes place in the facts that, in the former, $H$ has the $\Cc^1$ regularity 
w.r.t. $A$, while, in the latter, this regularity is absent and one replaces it by a much better 
regularity w.r.t. $\langle Q\rangle$. 
\end{rem}
In the previous results, we always chose $\Ic\subset ]0; k^2/4[$ (except in dimension $1$). What happens if 
we take $\Ic\subset ]k^2/4; +\infty[$? 
We shall give an answer in the case $\alpha =\beta=1$. Recall that, in this case, the form $[W_{11}, iA]$ 
is actually bounded on $\rL^2(\R^d)$. In the proof of Theorems~\ref{th:tal-A-Wigner-gj2} and~\ref{th:tal-Q-jm} above, we treated the contribution of $[W_{11}, iA]$ as a ``small'' pertubation of the dominant contribution given by $[H_0, iA]$. This was granted by the localisations in $H$ in the energy region $]0; k^2/4[$ 
(see Proposition~\ref{p:oscillations-energy1}). For the Mourre estimate \eqref{eq:esti-mourre-J}, the contribution of $[W_{11}, iA]$ then disappeared in the compact operator $K$. 
But \eqref{eq:esti-mourre-J} would still hold true if this contribution would be small compared to $[H_0, iA]=2H_0$. 
Roughly speaking, $2H_0$ is of the size of $2H$ and the latter is controlled by the considered interval $\Ic$. 
When the infimum of $\Ic$ is large enough compared to the size of $|w|$ in \eqref{eq:W} or, equivalently, 
when $|w|$ is small enough w.r.t. to this infimum, one can show that, for some compact operator $K_1$, 
\[\theta (H)\bigl(2H_0\, +\, [W_{\alpha\beta}, iA]\bigr)\theta (H)\ \geq \ \theta (H_0)^2H_0\, +\, K_1\, .\]
This is enough to replace the use of Proposition~\ref{p:oscillations-energy1} in the proof of the Mourre estimate 
(as pointed out in \cite{jm}, Remark 1.11) and also in the proof of Theorem~\ref{th:tal-A-Wigner-gj2}. 
In particular, the LAP \eqref{eq:tal-A} holds true on a compact interval $\Ic$ that is located at large enough energy and also holds true on any compact interval if $|w|$ is small enough, provided that $\Ic$ avoids the point 
spectrum of $H$. This was also obtained but in a different way in \cite{mu}. \\
What can we say if $d>1$, $\Ic\subset ]k^2/4; +\infty[$, and $|w|$ is large? One can show the LAP 
\eqref{eq:tal-Q} on $\Ic$ as stated in 
\begin{theorem}\label{th:tal-Q-m}\cite{mb}
Let $V=W_{11}+V_{sr}+V_{lr}$ and take a compact interval 
$\Ic$ included in $]k^2/4; +\infty[$. Then, for any $s>1/2$, \eqref{eq:tal-Q} holds true. 
\end{theorem}
\begin{rem}\label{r:tal-Q-m}
Again Theorem~\ref{th:tal-Q-m} remains valid if $V$ contains $H_0$-compact, local singularities. 
\end{rem}
One can prove Theorem~\ref{th:tal-Q-m} along the lines of the proof of Theorem~\ref{th:tal-Q-jm}. There is 
however an important change. In the present situation, we know that the form 
$\theta (H_0)[W_{11}, iA]\theta (H_0)$ does not extend to a compact form on $\rL^2(\R^d)$. But one can 
use the following 

\begin{proposition}\label{p:oscillations-energy1-bis}\cite{mb}. 
Let $\theta\in \Cc_c^\infty (\R;\C)$ with small enough compact support in $]k^2/4; +\infty[$. Then, for any $\epsilon >0$, there exist three bounded operators $B_\epsilon$, $B_\epsilon '$, and $C_\epsilon$ on $\rL^2(\R^d)$ such that 
\[\theta (H_0)\sin \bigl(k|Q|\bigr)\theta (H_0)\ =\ C_\epsilon \, +\, \langle Q\rangle^{-1}B_\epsilon\, +\, 
B_\epsilon '\langle Q\rangle^{-1}\, ,\]
$\|C_\epsilon\|\leq\epsilon$, and, as $\epsilon\to 0$, $\|B_\epsilon\|+\|B_\epsilon '\|=O(1)$. 
\end{proposition}  
We got rid of the contribution of $\theta (H_0)[W_{11}, iA]\theta (H_0)$ in the proof of 
Theorem~\ref{th:tal-Q-jm} by saying that it is compact or it ``contains'' decay in $\langle Q\rangle$. In the present situation, we use Proposition~\ref{p:oscillations-energy1-bis} to say that it is, up to some error that is compact or that ``contains'' decay in $\langle Q\rangle$, small compared to the main contribution, namely the one of $\theta (H_0)[H_0, iA]\theta (H_0)$, provided we take $\epsilon$ small enough. 

Before we end this section with a sketch of the proof of Proposition~\ref{p:oscillations-energy1-bis}, let us sum up the above results on the LAPs for Schr\"odinger operators in Figure~\ref{dessin:1}. Depending on the parameters $\alpha$ and $\beta$, that enter in the oscillating potential \eqref{eq:W}, we marked several regions. In the blue one, one can prove the LAP \eqref{eq:tal-A} without energy restriction in $]0; +\infty[$ by Mourre commutator method with $A$ as conjugate operator. Above the dark and red lines, the LAP \eqref{eq:tal-Q}, again without energy restriction in 
$]0; +\infty[$, is proven by one-dimensional arguments, provided that the longe range part of 
the potential is radial. In the green region, the LAP \eqref{eq:tal-Q} on $\Ic$ inside $]0; k^2/4[$ 
($\R\setminus\{k^2/4\}$, if $d=1$) is obtained by Theorem~\ref{th:tal-Q-jm}. On the point B, the LAP 
\eqref{eq:tal-A} is valid below $k^2/4$ (cf. Theorem~\ref{th:tal-A-Wigner-gj2}) and the LAP \eqref{eq:tal-Q} 
is granted above $k^2/4$ (cf. Theorem~\ref{th:tal-Q-m}). \\
Recently, Mourre's commutator theory for Schr\"odinger operator was developped with new conjugate operators (\cite{na,mar}). In particular, the LAP \eqref{eq:tal-Q} is shown in \cite{mar} to hold on $\Ic$ inside $]0; k^2/4[$ ($\R\setminus\{k^2/4\}$, if $d=1$) if $2\alpha +\beta >3$. Using an appropriate version of Mourre's commutator theory, one can even include negative values of $\beta$. This increases the blue region in Figure~\ref{dessin:1} but it still does not cover the green region.

\begin{figure}
\setlength{\unitlength}{1cm}
\begin{center}
\begin{picture}(8.5,6)(-0.5,-0.5)
\put(0,0){\vector(1,0){10}}
\put(0,0){\vector(0,1){5}}
\put(3,-0.1){\line(0,1){0.2}}
\put(-0.1,3){\line(1,0){0.2}}
\put(-0.1,2.4){\line(1,0){0.2}}
\put(-0.1,1.5){\line(1,0){0.2}}
\put(-0.1,2){\line(1,0){0.2}}
\put(-0.1,-0.5){0}
\put(5.9,-0.5){2}
\put(2.9,-0.5){1}
\put(-0.7,2.3){4/5}
\put(-0.7,1.9){2/3}
\put(-0.7,1.4){1/2}
\put(9.8,-0.5){$\alpha$}
\put(-0.4,2.9){1}
\put(-0.4,4.8){$\beta$}

\color{red}
\put(0,0){\line(1,1){2.97}}
\put(3,1.5){\line(1,0){1.5}}
\put(4.5,1.5){\line(1,-1){1.5}}
\put(2.2,1.7){red}
\put(1.5,0.8){red}
\put(4,0.8){red}
\color{green}
\put(3,1.5){\line(0,1){1.5}}
\put(3.2,1.6){green}
\color{blue}
\put(0.2,1){blue}
\put(1.4,4){blue}
\put(5.9,4){blue}
\put(5.9,1){blue}
\put(3,3){\line(1,-1){1.5}}
\color{black}
\put(0.1,2){\line(1,0){1.87}}
\put(2.9,2){\line(1,0){0.2}}
\put(3.2,1.9){A}
\put(2.6,2.9){B}
\put(3,2.4){\line(5,-2){1}}
\put(4,2){\line(1,0){6}}
\end{picture}
\end{center}
\caption{LAP. $\Vc=$ \textcolor{blue}{blue}\, $\cup$\, \textcolor{green}{green}.}\label{dessin:1}
\end{figure}

Let us now sketch a proof of Proposition~\ref{p:oscillations-energy1-bis}. It relies on a decomposition of the radial symmetric operator $\theta (H_0)\sin \bigl(k|Q|\bigr)\theta (H_0)$ in spherical harmonics and on the one-dimensional version of the interference phenomenon observed in Proposition~\ref{p:oscillations-energy1} (as mentioned in 
Remark~\ref{r:oscillations-energie1}). \\
As guiding ideas, we note, on one hand, that, above $k^2/2$ and in dimension $1$, $\theta (H_0)\sin \bigl(k|Q|\bigr)\theta (H_0)$ is compact (cf. Remark~\ref{r:oscillations-energie1}), and, on the other hand, this operator in arbitrary dimensions should behave like in dimension $1$, since it is radial. We also point out that the proof of Proposition~\ref{p:oscillations-energy1} in \cite{gj2} is based upon some appropriate pseudodifferential calculus (see Lemma 4.3, Lemma C.1, and Proposition A.1 in \cite{gj2}). \\
Using spherical coordinates, one can map $\rL^2(\R^d)$ by a unitary transformation to a direct sum 
\[\bigoplus _{\alpha\in\D}\, \rL^2(\R^+; dr)\otimes S_\alpha\, ,\]
where $\D=\{\ell -1+d/2; \ell\in\N\}$ and, for $\alpha\in\D$, $S_\alpha$ is a $(d-1)$-dimensional vector space over $\C$. Furthermore $H_0$ is unitary equivalent to a direct sum 
\[\bigoplus _{\alpha\in\D}\, h_\alpha^{>0}\, ,\]
where, for $\alpha\in\D$, $h_\alpha^{>0}:=-\partial _r^2+\alpha r^{-2}$. To get Proposition~\ref{p:oscillations-energy1-bis}, it suffices that, for all $\alpha\in\D$, Proposition~\ref{p:oscillations-energy1-bis} with $H_0$ replaced by $h_\alpha^{>0}$ 
and with a $O(1)$, that is uniform w.r.t. $\alpha$, holds true. \\
The latter result can be deduced from the same result but for the operators 
$h_\alpha:=-\partial _r^2+\alpha r^{-2}$, acting in $\rL^2(\R; dr)\otimes S_\alpha$. 
This can be seen using the restriction operator $R: \rL^2(\R; dr)\dans\rL^2(\R^+; dr)$
defined by $Rf=f_{|\R^+}$. Furthermore, since each $h_\alpha$ and $\sin (k|Q_r|)$ act trivially on $S_\alpha$, it suffices to prove the following:
\begin{proposition}\label{p:small-plus-compact}
Take $\alpha\in\D$. For any $\epsilon>0$, one can find an $\alpha$-independent function $\theta\in\Cc_c^\infty (\R;\C)$, with small enough support in $]k^2/4; +\infty[$, and bounded operators $B_\epsilon (\alpha )$, $B_\epsilon '(\alpha )$, and $C_\epsilon(\alpha )$, on $\rL^2(\R)$ such that, uniformly w.r.t. $\alpha$, $\|C_\epsilon(\alpha )\|\leq\epsilon$, $\|B_\epsilon(\alpha )\|+\|B_\epsilon '(\alpha )\|=O(1)$, and 
\begin{equation}\label{eq:oscillations-unif-alpha}
\theta (h_\alpha)\sin \bigl(k|Q_r|\bigr)\theta (h_\alpha)\ =\ C_\epsilon(\alpha ) \, +\, \langle Q_r\rangle^{-1}B_\epsilon(\alpha )\, +\, 
B_\epsilon '(\alpha )\langle Q_r\rangle^{-1}\, .
\end{equation}
Here $Q_r$ denotes the multiplication operator by the variable $r$ in $\rL^2(\R)$.
\end{proposition}
Since $\theta$ is compactly supported, we expect that the (possible) singularity at $r=0$ of 
$h_\alpha$ does not significantly contribute to $\theta (h_\alpha)$. Therefore we should be 
able to smooth it out by replacing $h_\alpha$ by the pseudodifferential operator $-\partial _r^2+(1-\chi_1)(r)\alpha r^{-2}$, where $\chi _1\in\Cc_c^\infty (\R;\C)$ and $\chi _1=1$ near $0$. The symbol of this pseudodifferential operator even belongs to the symbol class 
used to prove Proposition~\ref{p:oscillations-energy1}. But, to get an uniform bounds w.r.t 
$\alpha$ for the operators $B_\epsilon (\alpha )$ and $B_\epsilon '(\alpha )$, we need 
that this symbol stays in a bounded set in the symbol class, as $\alpha$ runs in $\D$. This is however not true. \\
To overcome this difficulty, we proceed as follows. Given $\epsilon>0$, we shall choose 
an appropriate function $\theta$ and some large enough $\kappa\geq 1$, and consider, 
for $\alpha\in\D$, the operators $h_{\alpha; \kappa}=p_{\alpha; \kappa}^w$, the Weyl 
quantization of the symbols $p_{\alpha; \kappa}$ given by, denoting by $\tau$ the dual variable of $r$, 
\[p_{\alpha; \kappa}(r; \tau)\ =\ \tau^2\, +\, \alpha r^{-2}\cdot\chi\bigl(\kappa^{-1}\alpha r^{-2}\bigr)\, ,\]
where $\chi\in\Cc_c^\infty (\R;\C)$ and $\chi =1$ near $0$. Since all the derivatives of 
the function $t\donne t^2\chi (\kappa^{-1}t)$ are bounded, $p_{\alpha; \kappa}$ 
actually stays in a bounded set in the previous symbol class, as $\alpha$ varies in $\D$. \\
Now, we can follow the proof of Proposition~\ref{p:oscillations-energy1} in \cite{gj2} to 
get, for each $\kappa\geq 1$, the result of Proposition~\ref{p:small-plus-compact} with $h_\alpha$ replaced by $h_{\alpha; \kappa}$ and $C_\epsilon(\alpha )$ replaced by $0$. We point out that $\theta$ is chosen such that, for $\tau\in\R$, $\theta (\tau^2)\theta ((\tau -k)^2)=0$. This requirement does not depend on $\alpha$. \\
To end the proof, we transfer the result for $h_{\alpha; \kappa}$ to $h_\alpha$ by pertubations. As a main step, we use the fact that $h_{\alpha; \kappa}$ and $h_\alpha$ 
differ in the region where $\alpha r^{-2}\geq \kappa$ and the positivity of $-\partial _r^2$, to get 
\[\theta (h_\alpha)\ =\ \theta _1(h_\alpha)\theta (h_{\alpha; \kappa})\, +\, \kappa^{-1}O(1)
\ =\ \theta (h_{\alpha; \kappa})\theta _1(h_\alpha)\, +\, \kappa^{-1}\tilde O(1)\, ,\]
where $\theta_1\in\Cc_c^\infty (\R;\C)$ satisfies $\theta\theta _1=\theta$, and 
$O(1)$ and $\tilde O(1)$ are uniform w.r.t. $\alpha$ and $\kappa$. Finally, we obtain \eqref{eq:oscillations-unif-alpha} with 
\[C_\epsilon (\alpha)\ =\ \theta (h_\alpha)\sin\bigl(k|Q_r|\bigr)\kappa^{-1}\tilde O(1)\, +\, \kappa^{-1}O(1)\sin\bigl(k|Q_r|\bigr)\theta (h_{\alpha; \kappa})\theta _1(h_\alpha)\]
and we choose $\kappa$ large enough to get the result of Proposition~\ref{p:small-plus-compact}.

  
%

\begin{thebibliography}{xxxxxx}  


%
\bibitem[ABG]{abg}W.O.\ Amrein, A.\ Boutet de Monvel, V.\ Georgescu: {\em
 $C_0$-groups, commutator methods and spectral theory of $N$-body   
hamiltonians.}, Birkh\"auser 1996.   
%
\bibitem[Be]{ben}M.\ Ben-Artzi: {\em Smooth Spectral Theory}, Operator Theory: Advances and Applications, 111 (2010), 
119-182.
%
\bibitem[BD]{bd}M.\ Ben-Artzi, A.\ Devinatz: {\em Spectral and
  scattering theory for the adiabatic oscillator and related
  potentials.} J. Maths. Phys. 111 (1979), p.\ 594-607.
%
\bibitem[BMP]{bmp} A.\ Boutet de Monvel, D.\ Manda, R.\ Purice: {\em Limiting absorption principle for the Dirac 
operator.}
Annales IHP, section A, tome {\bf 58}, 4, (1993), p. 413-431. 
%
\bibitem[Co]{co} M.\ Combescure:
{\em Spectral and scattering theory for a class of strongly oscillating potentials}, Comm. Math. Phys. 73, 43-62
(1980).
%
\bibitem[CG]{cg} M.\ Combescure, J.\ Ginibre:
{\em Spectral and scattering theory for the Schr\"odinger operator with strongly oscillating potentials}, Ann. IHP, section A, tome 24, $n^o$ 1
(1976), p. 17-30.
%
\bibitem[CHM]{chm}J.\ Cruz-Sampedro, I.\ Herbst, R.\ Martinez-Avenda\~{n}o: {\em   
Pertubations of the Wigner-Von Neumann potential leaving the embedded eigenvalue fixed.} 
Ann. H. Poincar\'e 3 (2002) 331-345.   
%
\bibitem[CFKS]{cfks}H.L.\ Cycon, R.G.\ Froese, W.\ Kirsch, B.\ Simon: {\em Schr\"odinger 
operators with applications to quantum mechanics and   global geometry.} Springer Verlag 1987.   
%
\bibitem[DMR]{dmr}A.\ Devinatz, R.\ Moeckel, P.\ Rejto: {\em A
    limiting absorption principle for Schr\"odinger operators with Von
    Neumann-Wigner potentials.} Int. Eq. and Op. Theory, vol.\  14 (1991).

%
\bibitem[DR1]{dr1}A.\ Devinatz, P.\ Rejto: {\em A limiting absorption principle for Schr\"odinger
operators with oscillating potentials. Part I.} J. Diff. Eq. , 49, pp. 29-84 (1983).
%
\bibitem[DR2]{dr2}A.\ Devinatz, P.\ Rejto: {\em A limiting absorption principle for Schr\"odinger
operators with oscillating potentials. Part II.} J. Diff. Eq. , 49, pp. 85-104 (1983).
%
\bibitem[FH]{fh} R.G.\ Froese, I.\ Herbst: {\em Exponential bounds and   
absence of positive eigenvalues for $N$-body Schr\"odinger operators.}   
Comm.\ Math.\ Phys.\ 87, 429-447 (1982).   
%
\bibitem[GGM]{ggm} V.\ Georgescu, C.\ G\'erard, J.S.\ M\o ller: {\em Commutators,
$C\sb 0$-semigroups and resolvent estimates}, J. Funct. Anal. {\bf 216}, no 2, p.\  303-361, 2004. 
%
\bibitem[G\'e]{ge} C.\ G\'erard: {\em A proof of the abstract limiting
absorption principle by energy estimates}, J. Funct. Anal. 254 (2008), no 11, 2707-2724. 
%
\bibitem[GJ1]{gj} S.\ Gol\'enia, Th.\ Jecko: {\em A new look at
  Mourre's commutator theory}, Compl. Ana. Op. Theory, Vol. 1, No. 3, 
p.\ 399-422, August 2007. Also available on ArXiv: ``  arXiv:math/0607275 [math.SP]  ''.
%
\bibitem[GJ2]{gj2} S.\ Gol\'enia, Th.\ Jecko: {\em Weighted Mourre's commutator theory, application
to Schr\"odinger operators with oscillating potential}, J. Oper. Theory 70:1 (2013), 109-144.
Also available on ArXiv: `` arXiv:1012.0705 [math.SP]  ''.
%
\bibitem[HS]{hs} B.\ Helffer, J.\ Sj\"{o}strand: {\em Op\'erateurs de  
Schr\"{o}dinger avec champs magn\'etiques faibles et constants.}   
Expos\'e No.\ XII, S\'eminaire EDP, f\'evrier 1989, Ecole Polytechnique.  
%
\bibitem[K]{ki} A.\ Kiselev: \emph{Imbedded singular continuous
    spectrum for Schr\"odinger operators},  J.\ Amer.\ Math.\ Soc.\ 18
  (2005), no.\ 3, 571–603. 
%
\bibitem[JM]{jm}Th.\ Jecko, A.\ Mbarek: {\em Limiting absorption principle for Schr\"odinger operators with oscillating potential.} 
Documenta Mathematica\ 22 (2017), 727-776. 
%
\bibitem[JMP]{jmp}A.\ Jensen, E.\ Mourre, P.\ Perry: {\em Multiple
  commutator  estimates and resolvent smoothness in quantum scattering
 theory.} Ann.\  Inst.\ H. Poincar\'e vol.\ 41, no 2, 1984, p.\
  207-225. 
%
\bibitem[LNS]{lns}A.\ Laptev, D.\ Naboko, O.\ Safranov: {\em Absolutely continuous spectrum of Schr\"odinger operators 
with slowly decaying and oscillating potentials.} Comm. Math. Phys. {\bf 253}, (2005), p. 611-631. 
%
\bibitem[La1]{l1} R.B. Lavine: {\em Commutators and scattering theory. I. Repulsive interactions.}
Comm. Math. Phys. {\bf 20}, 301-323, (1971). 
%
\bibitem[La2]{l2} R.B. Lavine: {\em Commutators and scattering theory. II. A class of one body problems.}
Indiana Univ. Math. J., Vol. {\bf 21}, No. {\bf 7}, 643-656, (1972). 
%
\bibitem[LS]{ls} J.\ L\"orinczi, I.\ Sasaki: {\em Embedded eigenvalues and Neumann-Wigner potentials for relativistic 
Schr\"odinger operators.} J. Funct. Anal. 273 (2017), no. 4, 1548–1575. 
%
\bibitem[Man]{man}M.\ Mandich: {\em The limiting absorption principle for the discrete Wigner-Von Neumann operator.} 
Preprint. Available on ``http://arxiv.org/abs/1605.00879  '' and ``   https://hal.archives-ouvertes.fr/hal-01297239  ''.
%
\bibitem[Mar]{mar}A.\ Martin: {\em On the limiting absorption principle for Schr\"odinger Hamiltonians.} Preprint. Available on 
``  http://arxiv.org/abs/1510.03543  `` and ``  https://hal.archives-ouvertes.fr/hal-01214523  ''. 
%
\bibitem[Mb]{mb}A.\ Mbarek: {\em \'Etude du théorème d'absorption limite pour des opérateurs de Schr\"odinger et Dirac avec un potentiel oscillant.}. Phd thesis, 2017. 
%
\bibitem[MU]{mu}K.\ Mochizuki, J.\ Uchiyama: {\em Radiation conditions
    and spectral  theory for 2-body Schr\"odinger operators with
    ``oscillating'' long-range potentials  I: the principle of
    limiting absorption.} J.\ Math.\ Kyoto Univ.\ {\bf 18}, 2,  
377--408, 1978.
%
\bibitem[Mo]{m}E.\ Mourre: {\em Absence of singular continuous spectrum for   
certain self-adjoint operators.} Comm.\ in Math.\ Phys.\ {\bf 78}, 391--408, 1981.  
%
\bibitem[Nab]{n}S.\ N.\ Naboko: {\em Dense point spectra of Schr\"odinger and Dirac operators.} Theor. Math. Phys. {\bf 68},
(1986), No. 1, 646--653. 
%
\bibitem[Nak]{na}S.\ Nakamura: {\em A remark on the Mourre theory for two body Schr\"odinger operators.} J. Spectral Theory {\bf 4},
(2015), No. 3, 613--619. 
%

 
%
\bibitem[RS3]{rs3}M.\ Reed and B.\ Simon : {\em Methods of Modern
    Mathematical Physics, Tome III: Scattering theory.} Academic Press.   
%
\bibitem[RS4]{rs4}M.\ Reed, B.\ Simon: {\em Methods of Modern
    Mathematical  Physics, Tome IV: Analysis of operators.} Academic Press.  
%
\bibitem[Re]{rem} C.\ Remling: \emph{The absolutely continuous spectrum
    of one-dimensional Schr\"odinger operators with decaying
    potentials}, Comm.\ Math.\ Phys.\ {\bf 193} (1998), 151--170. 

\bibitem[RT1]{ret1}P.\ Rejto, M.\ Taboada: {\em A limiting absorption
    principle for Schr\"odinger operators with generalized Von
    Neumann-Wigner potentials I. Construction of approximate phase.}
  J.\ Math. Anal. and Appl. 208, p.\ 85--108 (1997). 
%
\bibitem[RT2]{ret2}P.\ Rejto, M.\ Taboada: {\em A limiting
    absorption principle for Schr\"odinger operators with generalized
    Von Neumann-Wigner potentials II. The proof. } J. Math.
  Anal. and Appl. 208, p.\ 311-336 (1997). 
%
\bibitem[RUU]{ruu}S.\ Richard, J.\ Uchiyama, T.\ Umeda: {\em Schr\"odinger operators with n positive eigenvalues: an 
explicit construction involving complex-valued potentials.} Proc. Japan Acad. Ser. A Math. Sci.\  92 (2016), no. 1, 7–12. . 
%
\bibitem[Sa]{sa}J.\ Sahbani: {\em  The conjugate operator method for locally 
  regular Hamiltonians.} J.\ Oper.\ Theory 38, No.\ 2, 297--322
  (1997).
  
 \bibitem[Sc]{sc}K.M.\ Schmidt: {\em Dense point spectrum for the one-dimensional Dirac operator with an electrostatic potential.} Proc. Royal Soc. Edinburgh 126A, 1087-1096, 1996. 
%
\bibitem[Si1]{si1}B.\ Simon: {\em  On positive eigenvalues of one body
  Schr\"odinger operators.} Comm.\ Pure Appl.\ Math.\ 22, 531--538 (1969).
%
\bibitem[Si2]{si2}B.\ Simon: {\em Some Schr\"odinger operators with dense point spectrum.} Proc.\ Amer.\ Math.\ Soc.\ {\bf 125}, 
1, 203--208 (1997).
%
\bibitem[T]{th}B.\ Thaller: {\em The Dirac Equation.} Springer-Verlag Berlin Heidelberg New York, 1992. 
%
\bibitem[VW]{vw}J.\ Von Neumann, E.\ Wigner: {\em  \"Uber merkw\"urdige diskrete Eigenwerte.} Z.\ Phys.\ {\bf 30}, 
465--467 (1929).
%
\bibitem[W]{w}D.\ A.\ W.\ White: {\em  Schr\"odinger operators with rapidly oscillating central potentials.} Trans. 
AMS {\bf 275}, 2, February 1983. 
%
\bibitem[Y]{y}D.\ D.\ R.\ Yafaev: {\em Mathematical Scattering theory. General theory.} AMS (1992). 
%
\end{thebibliography}
\end{document}